\documentclass[prd,twocolumn,tightenlines,floatfix]{revtex4}
\usepackage{graphicx,subfigure}
\usepackage{amsmath,amssymb}
\usepackage{float}
\usepackage{xcolor}
\usepackage{bm}
\usepackage{lineno}

\newcommand{\be}{\begin{eqnarray}}
\newcommand{\ee}{\end{eqnarray}}

\def\slashchar#1{\setbox0=\hbox{$#1$}           
   \dimen0=\wd0                                 
   \setbox1=\hbox{/} \dimen1=\wd1               
  \ifdim\dimen0>\dimen1                        
 \rlap{\hbox to \dimen0{\hfil/\hfil}}      
  #1                                        
 \else                                        
    \rlap{\hbox to \dimen1{\hfil$#1$\hfil}}   
    /                                         
 \fi}                                         %

\begin{document}

\title{Towards a semi-classical description of QCD vacuum around $T_c$}

\author{ Rasmus N. Larsen}
\email{rlarsen@bnl.gov}
\affiliation{Physics Department, Brookhaven National Laboratory, Upton NY 11973}
\author{Sayantan Sharma}
\email{sayantans@imsc.res.in}
\affiliation{The Institute of Mathematical Sciences, HBNI, Chennai 600113} 
\author{ Edward  Shuryak }
\email{edward.shuryak@stonybrook.edu}
\affiliation{ Department of Physics and Astronomy, Stony Brook University, 
Stony Brook, NY 11794 }

\begin{abstract}
We study the vacuum topology of 2+1 flavor QCD above the chiral crossover 
transition, at $T \lesssim 1.2~T_c$,  on lattices of size $32^3\times 8$. 
Since overlap fermions have exact chiral symmetry and an index theorem even 
on a finite lattice, we use them to detect the topological content of gauge 
fields generated using domain wall fermion discretization for quarks.  We 
further use different periodicity phases along the temporal direction 
for the valence overlap quarks, which allows us to probe different topological 
structures present in the gauge field ensembles, through its zero modes.  This 
procedure provides strong evidences that fermion zero-modes can be quantitatively 
understood to arise due to different species of instanton-dyons. We estimate 
their relative abundances from the Dirac-eigenvalue density and resolve 
the so called ``topological clusters" via multi-parameter fits to their density, 
providing therefore an understanding of the interactions between instanton-dyons. 
The typical separation between dyons we obtain, is $\sim 0.3$ fm. 
Surprisingly, it emerges out from this study that a semi-classical description 
of the fermionic zero modes in the QCD vacuum is quite accurate just above $T_c$. 
\end{abstract}
\maketitle


\section{Introduction}
With the increasing sophistication of non-perturbative lattice gauge theory 
techniques, we now  know that the (nearly exact) chiral symmetry in 
2+1 flavor Quantum Chromodynamics (QCD) is restored via a continuous and 
smooth crossover transition ~\cite{Aoki:2006we,Bazavov:2011nk,Bhattacharya:2014ara}. 
The chiral crossover transition temperature was recently quantified to an unprecedented 
accuracy \cite{Bazavov:2018mes} and is $T_c=156.5\pm1.5$ MeV. The observables sensitive 
to chiral symmetry breaking and confinement show consistent behavior in the crossover transition, implying a simultaneous occurrence of these two phenomena in QCD with quarks 
in the fundamental representation of the gauge group~\cite{Bazavov:2016uvm}. The question 
that still remain unresolved is the precise topological identity of the microscopic degrees 
of freedom responsible for these two physically distinct phenomena in QCD.

With the motivation of explaining confinement in QCD, Polyakov introduced 
instantons. In the  2+1 D  case he showed in his seminal work~\cite{Polyakov:1976fu} 
that a gas of instantons (monopoles in this setting) produces the famous 
confining potential. However in the (most important) 3+1D case, the ensemble of 
instantons were not confining due to the fact that their interaction is 
not long-range. The instanton liquid model (ILM) \cite{Shuryak:1981ff} 
used the instanton density and the average radius of the instantons as inputs. 
With their values known, one can show that the instanton-induced t'Hooft 
Lagrangian is strong enough to break $SU_A(N_f)$ chiral symmetry, like it was 
the case for the Nambu-Jona-Lasinio (NJL) model~\cite{Nambu:1961tp}. 
Further studies of interacting instanton ensembles predicted different 
hadronic correlation functions within ILM, which are in fairly good 
agreement with the results from QCD inspired models and lattice~\cite{Schafer:1996wv}.
This model explained not only the $SU_A(N_f)$ chiral symmetry breaking and pions, 
but also large violations of the $U_A(1)$ symmetry, e.g. the large mass of the 
$\eta'$ meson.  However confinement could not be explained within this framework.

The (renormalized) Polyakov loop is traditionally used as the order parameter 
of the confinement transition in pure gauge theory. In order to understand both 
deconfinement and chiral symmetry breaking, one needs to find topological solitons 
sensitive to the eigenvalues of the Polyakov loop. At finite temperatures, such 
solitons with a non-trivial holonomy have been shown to exist in~
\cite{Lee:1998vu,Kraan:1998sn,Lee:1998bb,Kraan:1998pm}: they are called 
instanton-dyons or instanton-monopoles. For simplicity, we henceforth 
refer to them as ``dyons". 
Thermodynamic ensembles of dyons were studied analytically~
\cite{Liu:2015ufa,Liu:2015jsa,Liu:2016thw,Liu:2016yij} via mean-field 
approximation, as well as numerically~
\cite{Larsen:2015vaa,Larsen:2015tso,Larsen:2016fvs}.
These works could explain (near simultaneous) occurrence of both 
deconfinement and chiral phase transitions in QCD-like theories. 
Unlike instantons, the instanton-dyons carry not only topological 
but also electric charges, thus back-interacting with the Polyakov 
loop inducing confinement~\cite{GarciaPerez:1993ab,GonzalezArroyo:1996jp}. 
They also carry magnetic charges and therefore Dirac strings, 
which explains why their topological charges are not quantized 
as integers.  

Whether this is indeed the right explanation or not, can be directly 
tested using  lattice gauge-field configurations. One way to do so is 
by using over-improved cooling technique~\cite{GarciaPerez:1999hs,
Bornyakov:2013iva}. The other one, which we use, is via studying the 
properties of the fermionic zero and near-zero modes, with modified 
temporal periodicity phases as a tool~\cite{Gattringer:2002wh}. 
Studies along these lines, have been performed by Mueller-Preussker, 
Ilgenfritz and collaborators, see for e.g.~\cite{Bornyakov:2015xao,Bornyakov:2016ccq} 
which provided tantalizing evidences for the presence of dyons also 
in finite temperature QCD.

The aim of this work is to provide quantitative description of 
these Dirac eigenstates in terms of superposition of the dyons. 
Using sophisticated lattice techniques we aim to isolate and 
identify different species of dyons and understand the nature 
of interactions between them. For the first time, fermions with 
exact chiral symmetry on the lattice, the so-called overlap fermions, 
are used for this aim. Our initial results have been published in 
Ref.~\cite{Larsen:2018crg}. Even though computationally expensive,  
such fermions are ideally suited to probe the topological content 
of the gauge configurations, which, in our case, are  generated with 
the domain-wall fermion discretization.
By varying the temporal periodicity phases of the (valence) 
fermions, we track all fermionic zero modes. We show that 
not only the index theorem works as expected, but the objects 
responsible for the zero modes are indeed nothing but different 
species of dyons. Further nontrivial tests of this picture 
is a comparison of the quark zero-modes~\emph{profiles}, obtained 
numerically from lattice, to the known analytic solutions within 
the semi-classical theory of dyons. While these profiles are not 
in general topologically protected, we still find a surprisingly 
good agreement between the analytic (semi-classical) and lattice 
profiles. Continuing forward from our preliminary findings 
reported in Ref.~\cite{Larsen:2018crg}, we describe here our
findings in a more extensive form, with several new results. 

The main conclusion of our work is that a semi-classical description 
of the fermionic zero-modes of QCD in terms of an ensemble of dyons 
is quite accurate, just above $T_c$. In a more broader context, let 
us note that recent lattice measurements of the topological 
susceptibility in QCD~\cite{Petreczky:2016vrs,Frison:2016vuc,
Borsanyi:2016ksw,Burger:2018fvb,Bonati:2018blm} were related to 
a semi-classical description, in terms of a dilute gas of instantons
~\cite{Gross:1980br} only at fairly high temperatures, beyond $3~T_c$. 
Our study, focusing on a different temperature regime just above $T_c$, 
sheds light on the instanton \emph{ionization} phenomenon.

The paper is organized as follows. In section~\ref{sec_comp} we describe 
the basic computational techniques and resources used in this work.
Then, in section~\ref{sec_dyons} we review the known expressions 
for the corresponding solutions of the Dirac equation (the zero modes) 
in the dyon background.  Next section~\ref{sec_results} contains our 
main results, presented in a series of sub-sections. We show our 
lattice results for the quark zero-modes as a function of temperature 
for different choices of the temporal periodicity phases for the valence 
fermions. We qualitatively describe the Dirac zero-modes obtained from 
lattice calculations, emphasizing the evidences that those can be explained 
in terms of dyons. For overlapping as well as well-separated dyons we show that 
the lattice fermionic density  agrees well with those calculated in the semi-classical 
theory of dyons. For the first time, we calculate distribution of the dyons in sizes 
and locations,   as a function of temperature. We finally conclude with possible 
applications of our results for a more comprehensive understanding of the crossover 
transition of QCD in terms of its topological content.

\section{Computational techniques} 
\label{sec_comp}

As already mentioned in the Introduction, our aim is to use 
lattice fermions with exact chiral symmetry and index theorem 
on the lattice for understanding the topology in hot QCD. This 
is achieved in two ways. 
First of all, the lattice configurations for $2+1$-flavor QCD  
at finite temperatures has nearly exact chiral symmetry. These 
were generated by the HotQCD collaboration using M\"{o}bious 
domain-wall discretization for fermions~\cite{Kaplan:1992bt,
Brower:2012vk} and Iwasaki gauge  action, previously used for studies 
of the chiral crossover transition temperature $T_c$ and $U_A(1)$ 
breaking in QCD~\cite{Bhattacharya:2014ara}. 
 
The lattices have spatial dimensions $N_s=32$, with $N_\tau= 8$  
temporal sites and $N_5=16$ sites along the fifth dimension. For 
this  extent  of $N_5$, the residual chiral symmetry breaking due 
to this specific fermion discretization on the lattice, is of the 
order of $\sim 2\times 10^{-3}$. The pseudo-critical temperature 
measured using these configurations is $T_c=155(9)$ MeV~
\cite{Bhattacharya:2014ara}. We have selected four temperatures 
$T/T_c=1.0, 1.08, 1.1, 1.2$ for our study.  While these may appear 
very close to each other, this selection is in fact rather wide, 
because the relevant quantity for dyons, the average Polyakov loop, 
varies significantly between near zero and $\mathcal{O}(1)$ values. 
Therefore, our choice is well suited for studies of physical phenomena 
that depend crucially on the non-trivial holonomy.  A more detailed 
discussion of the holonomy of these analyzed configurations can be 
found in Sec.~\ref{sec:hol}.

Some configurations studied were selected to have 
total topological charge $|Q|= 1$, and respectively a single zero 
mode. We found this way is the cleanest set-up to identify and study 
exactly the zero modes. We have also studied configurations with $|Q|=0,2,3$  
to check the robustness of our results. The details of the configurations 
used in our analysis is summarized in Table~\ref{tab:config}. 

\begin{table}
\centering
\begin{tabular}{  | c | c | c | c | c |}
\hline
\hline
    $T/T_c$ & $N_s$ & $N_\tau$ & \# Configurations & $\vert Q_{top}\vert$\\ \hline
  1.00 & 32 & 8 & 15 & 1\\
   \hline 
    1.08 & 32 & 8 & 15 & 1\\
   \hline 
    1.1 & 32 & 8 & 33 & 0,1,2\\
   \hline 
    1.2 & 32 & 8 & 30 & 0,1\\
   \hline 
   \hline
 \end{tabular}
 \caption{Details of the QCD gauge ensembles used in this work.}
 \label{tab:config}
 \end{table}

Unfortunately, the domain wall fermions used in generating the 
configurations, turned out to be not precise enough for our purposes. 
Due to small but finite residual chiral symmetry breaking, the domain 
wall fermions used in simulation do not satisfy an exact index theorem 
on the lattice, neither do they have well defined zero-modes. Therefore 
we have instead used another formulation of lattice fermions, known as 
the overlap fermions~\cite{Neuberger:1998my}, 
\begin{eqnarray}
D_{ov} = 1-\gamma_5 sign(H_W)~,H_W = \gamma _5 (M - a D_W)
\end{eqnarray}
as the so called \emph{valence fermion operator}, where $D_W$ is the 
Wilson-Dirac operator for massless quarks in four dimensions and 
$M$ is the domain wall height which is chosen to be $M=1.8\in[0,2)$ 
to simulate one quark flavor on the lattice. Owing to the fact that 
the overlap operator has an exact index theorem, even for a finite 
lattice spacing~\cite{Hasenfratz:1998ri}, its zero modes must be 
related to topological objects resigning in the underlying gauge 
configurations. 

We have implemented the overlap operator on the lattice by explicitly evaluating 
the \emph{sign}-function via deflation on the eigen-space of $H_W^2$ consisting 
of the lowest $25$ eigenvectors and approximating on the rest of the eigen-space 
by a Zolotarev Rational Polynomial with $25$ terms. We have ensured that the sign 
function is realized as precise as $\sim 10^{-9}$. Furthermore we have checked 
that  for each gauge configuration the resulting overlap-Dirac operator satisfies 
the Ginsparg-Wilson relation~\cite{Ginsparg:1981bj},
 \begin{eqnarray}
\gamma _5 D_{ov}^{-1} + D_{ov}^{-1} \gamma _5 &=&  \gamma _5 
\end{eqnarray}  
to a precision of $10^{-9}$ or better.

\section{From Calorons to Instanton-dyons and  their Dirac zero modes } 
\label{sec_dyons} 
The term \emph{caloron} is used for  the finite temperature classical 
solutions of gauge fields~\cite{Harrington:1978ve} with unit topological 
charge and \emph{trivial} Polyakov loop vacuum expectation value (VEV). 
A crucial new step made in~\cite{Kraan:1998sn,Lee:1998bb} was to generalize 
it to the case in which the Polyakov loop has a \emph{nontrivial} VEV. 
In the relevant case of  the $SU(3)$ gauge theory, it can be represented 
by three phases $\mu_{1,2,3}$, called \emph{holonomies}, such that the 
average value of the Polyakov loop is 
$\mathcal P= \frac{1}{3} Tr[\exp[i\text{Diag}(\mu _1,\mu _2,\mu _3)]]$.
These three phases can be measured on lattice configurations, both 
locally, and as an average over the entire spacetime volume.

\begin{figure}[]
\begin{center}
\includegraphics[width=7cm]{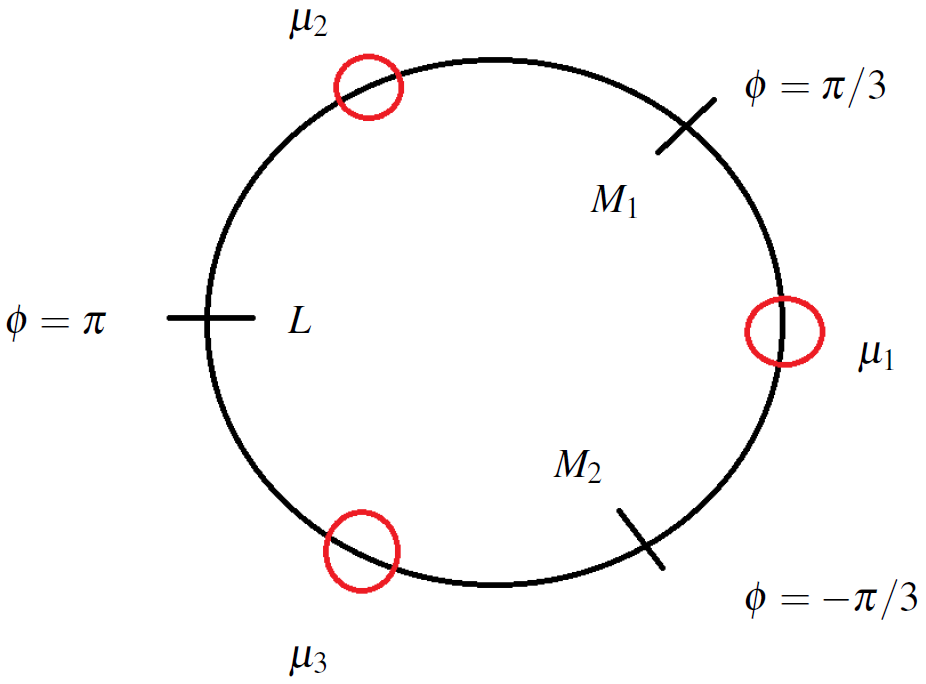}
\caption{A representation of the different angles of the Polyakov loop 
$\mu _i$ on unit a circle and the three phases corresponding to the 
fermionic boundary condition along the temporal direction, which we use 
in this paper. $M_1$, $M_2$ and $L$ indicate the name of the dyon in the 
specific segment of the circle. The $\mu_i$'s have been chosen to represent 
a typical configuration in the confined phase of QCD.}
\label{fig:circle}
\end{center}
\end{figure}

The classical gauge field solutions of Yang-Mills equation with 
arbitrary $\mu_{1,2,3}$,  based on a combination of Nahm 
transformation~\cite{Nahm:1979yw} and the Atiyah-Hitchin-
Drinfeld-Manin (ADHM) construction~\cite{Atiyah:1978ri}, 
is rather involved. Let us just mention that this solution 
has twelve parameters, out of which nine of them are the spatial 
locations of the three distinct dyons,  with the additional three 
parameters being the $U(1)$ phases. If the locations of dyons are 
well separated in space, then for an appropriate gauge choice 
(the hedgehog gauge), one finds that these dyons are basically 
monopoles, with the gauge potential $A_0$ as a scalar field.
It is instructive to put these phases $\mu_{1,2,3}$ on a unit 
circle, as shown in Figure~\ref{fig:circle} by small red circles. 
One of the phases $\mu_1$ of the Polyakov loop is real, and two 
others are complex conjugates of each other. The explicit values of 
the phases shown in Figure~\ref{fig:circle} is a typical representation 
in the confined phase of QCD. The actions for each of these dyon types are 
proportional to the lengths of arcs (segments)  between two successive 
phases on the circle, e.g. for $M_1$-dyon the action is proportional 
to $\mu_2-\mu_1$. Since the three arcs together make a complete circle, 
the sum of the action of the three individual dyons is always equal to 
$8\pi^2/g^2$, the action of the caloron.  Since the solutions are 
(anti)self-dual, the action and the topological charges are 
(up to a sign) proportional to each other. Dyons thus have  
non-integer topological charges. 

And yet, from the index theorem, the number of Dirac zero modes can 
only be an integer. This would naively imply that by counting the 
number of zero modes of the QCD Dirac operator, we will always get 
an integer topological charge and can never identify the individual 
dyons that constitute a caloron. To resolve this dilemma we recall 
that in QCD at finite temperature, all quark fields acquire a phase 
$\phi_u=\phi_d=\phi_s=\pi$ when traversed once along the Euclidean 
temporal direction. This is the so-called anti-periodic boundary 
condition due to the Grassman nature of the quarks. However, one 
can assign any arbitrary periodicity phase $\phi$ one wants, to 
the valence quarks, if we use the valence fermion zero modes merely 
as a tool, or a filter, to identify the topological content of the 
gauge configurations.  
The identification rules are now simple;  depending on which segment 
of the phase circle the $\phi$ lies in, the dyon corresponding to that 
segment produces a normalizable zero-mode of the Dirac operator. For
the other dyon backgrounds, the corresponding Dirac operator with phase 
$\phi$ will have non-normalizable solutions, which are not physical. 
In this work, we have selected three possible phases for the valence 
overlap fermions, $\phi=\pi,\pm \pi/3$ as shown in Figure~\ref{fig:circle}. 
Since these values are placed in segments corresponding to $L,M_1,M_2$-type 
dyons, these three choices would let us identify all three types of dyons via 
its zero modes. Obviously this method of detecting dyons by tuning the periodicity 
phase of the valence Dirac operator works best if the dyons are well separated 
i.e., when the calorons are \emph{ionized} into its constituent dyons. In this 
work we will apply this method not only to well-separated dyons, but also
to \emph{topological clusters},  made of considerably overlapping dyons, 
see  Sec.~\ref{sec_results}.

As it will be clear from the analysis to follow, we will fit the 
shape of the topological clusters obtained from the lattice with 
the Dirac zero modes of overlapping dyons derived analytically in 
the semi-classical approximation. The details of the fitting procedure 
is explained in Appendix A, where we have numerically calculated the 
zero-mode density used in the fit, which is outlined in Appendix B.  
Here we only review the basic theoretical foundations for the derivation 
of fermion zero modes on dyon backgrounds. It has been derived earlier
~\cite{GarciaPerez:1999ux} that the fermion zero-mode \emph{density} can 
be related to the Laplacian of a certain Green's function (diagonal part) 
which is defined on the circle of phases of the Polyakov loop~\cite{Chernodub:1999wg}. 
The fermion zero-mode density $\rho(x)$ at any 4D spacetime point $x$ 
is derived to be,
\begin{equation}
\label{eqn:zmode1}
\rho(x) = -\frac{1}{  4\pi^2}\partial _{\mu} ^2 f_x( \phi, \phi)~,
\end{equation}
where the Green's function $f_x( \phi, \phi')$ itself is the solution of the 
differential equation,
\begin{equation}
\label{eqn:zmode2}
 \left( D_{\phi}^2 
+r^2(x,\phi)+\sum _{m=1} ^3 \delta _m(\phi) \right)
f_x(\mathbf \phi,\mathbf \phi') =\delta(\mathbf \phi-\mathbf \phi')~. 
\end{equation}
Here the functional derivative 
$D_{\mathbf \phi} =\frac{1}{i}\partial _{\mathbf \phi} -\tau $ is the only 
place where Euclidean time $\tau$ appears. The \emph{sources} correspond 
to the locations of the phases $\mu_m$ of the Polyakov loop and are 
represented as, 
$\delta _m(\mathbf \phi)=\delta (\mathbf \phi-\mu_m)|x_m-x_{m+1}|/(2\pi)$
with the corresponding strengths proportional to distances between the 
centers of the $m$-th and the $(m+1)$-th dyons, which are given as 
$x_m,x_{m+1}$ respectively where $m=1,2,3$. The $x$-dependent 
\emph{potentials} are 
$r^2(x,\mathbf \phi)=r_m ^2(x),\mathbf\phi\in [\mu_m, \mu_{m+1}]$  
defined differently in corresponding segments since $r_m(x)$ 
is the distance between the observation point $x$ and the center 
of the $m$-th dyon. Till recently the analytic solution of this 
equation on a torus was only available for the $SU(2)$ gauge group~
\cite{GarciaPerez:1999ux}, the case with only two species 
($L$ and $M$) of dyons. Since here we are working with the 
$SU(3)$ gauge group, we solve Eq.~\ref{eqn:zmode2} numerically, 
details of which are outlined in Appendix B. It is only 
during finalizing our manuscript we learned that an analytic 
solution for the gauge potentials in a general $SU(N_c)$ gauge 
group with a non-trivial holonomy on a torus has been worked 
out in Ref.~\cite{Gonzalez-Arroyo:2019wpu}. It would be now 
possible to compute the fermion zero-modes on such backgrounds.

We will henceforth use the numerical solution to fit it to the 
topological clusters obtained from lattice simulations, and will 
successfully identify and describe quantitatively different clusters. 
In particular, the spacetime locations of all dyons, as well as the 
Polyakov loop values in terms of its phases $\mu_i$ will be determined
in each case. This will enable us to distinguish between overlapping 
set of all three  species of dyons on one hand, and a caloron with a 
trivial holonomy on the other. 

We have implemented the quark periodicity phases $\phi$ along the temporal 
direction for valence overlap Dirac operator by setting the gauge links 
along the temporal boundary to 
$U_4(\beta+1/T)\rightarrow \rm{e}^{i\phi} U_4(\beta)$,
corresponding to three different choices of $\phi=-\pi/3,\pi/3,\pi$, 
the last one corresponding to the usual anti-periodic boundary condition.  
As evident from Figure~\ref{fig:circle}, these choices of $\phi$ ensure 
that the zero modes are equidistant from each other and thus not too close 
to each $\mu_m$, where its wavefunction density has a discontinuous jump 
from one dyon sector to the other. 
For these three choices of $\phi$ we have then calculated, at each temperature,  
the first $6$ eigenvalues and eigenvectors of the valence overlap Dirac operator,   
on the (domain wall) configurations using the Kalkreuter-Simma Ritz algorithm~
\cite{Kalkreuter:1995mm}. At the two highest temperatures $T/T_c=1.1,1.2$ we 
have also calculated the first $30$ eigenmodes of the QCD Dirac operator to 
understand the dependence of the eigenvalue spectrum on the quark periodicity 
phases $\phi$. Two gauge-invariant  observables for each eigenstate are its 
density $\rho(x)$ and the chiral density $\rho_5(x)$, defined at each 
spacetime point $x$ as,
\begin{eqnarray}
\nonumber
\rho (x) = \sum_{a=1}^3\sum_{i=1}^{4}\psi^\dagger_{a,i} \psi_{a,i} ~,~
\rho _5 (x) =\sum_{a=1}^3\sum_{i=1}^{4}\psi^\dagger_{a,i}\gamma_{5(i,j)}\psi_{a,j}
\end{eqnarray}
Studying the spacetime profiles of the zero modes and tracking their change 
in spatial positions for different boundary conditions on the lattice, we 
identify the presence of calorons or its substructures, the dyons. The 
eigenvalue method is numerically unambiguous compared to cooling or Wilson flow 
techniques, since the lowest eigenmodes are automatically separated from 
the ultra-violet modes, making them insensitive to large ultra-violet 
fluctuations in the gauge field. This therefore, allows us to study 
the infrared  modes of the QCD spectrum in a non-invasive manner.

\section{Results} 
\label{sec_results} 

\subsection{Properties of quark zero-modes near $T_c$}

Let us start this  section with the following question:
are the dyons separate entities, appearing at different locations, or
are they simply parts of certain multi-dyon clusters, e.g. the original 
calorons?  In the latter case, the positions of the zero-mode density 
peaks will remain localized at the same  point, whereas in the former 
case, they will be at different locations, as the valence quark periodicity 
phase $\phi$ is changed.

\begin{figure}[]
\begin{center}
\includegraphics[width=6cm]{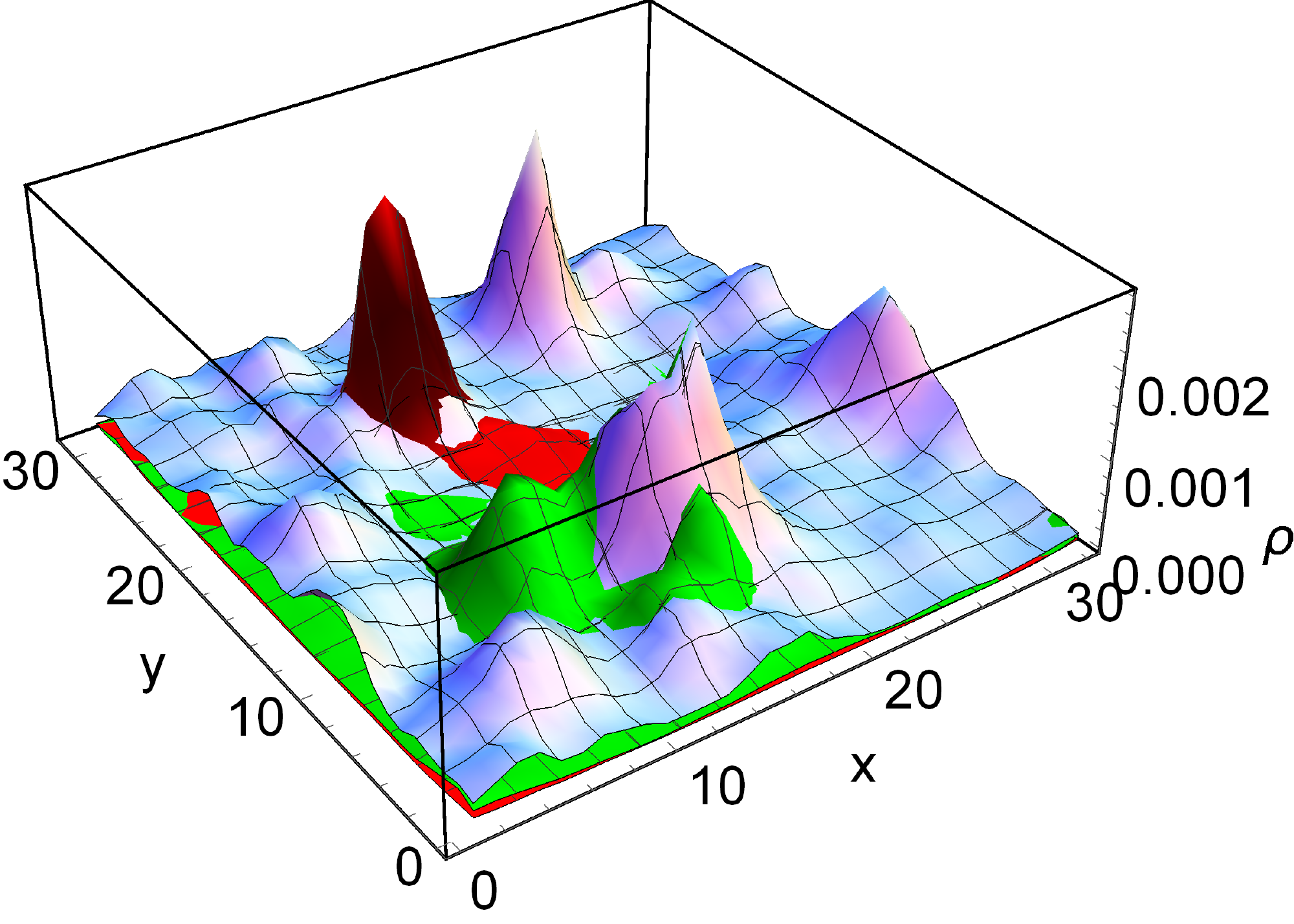}
\includegraphics[width=6cm]{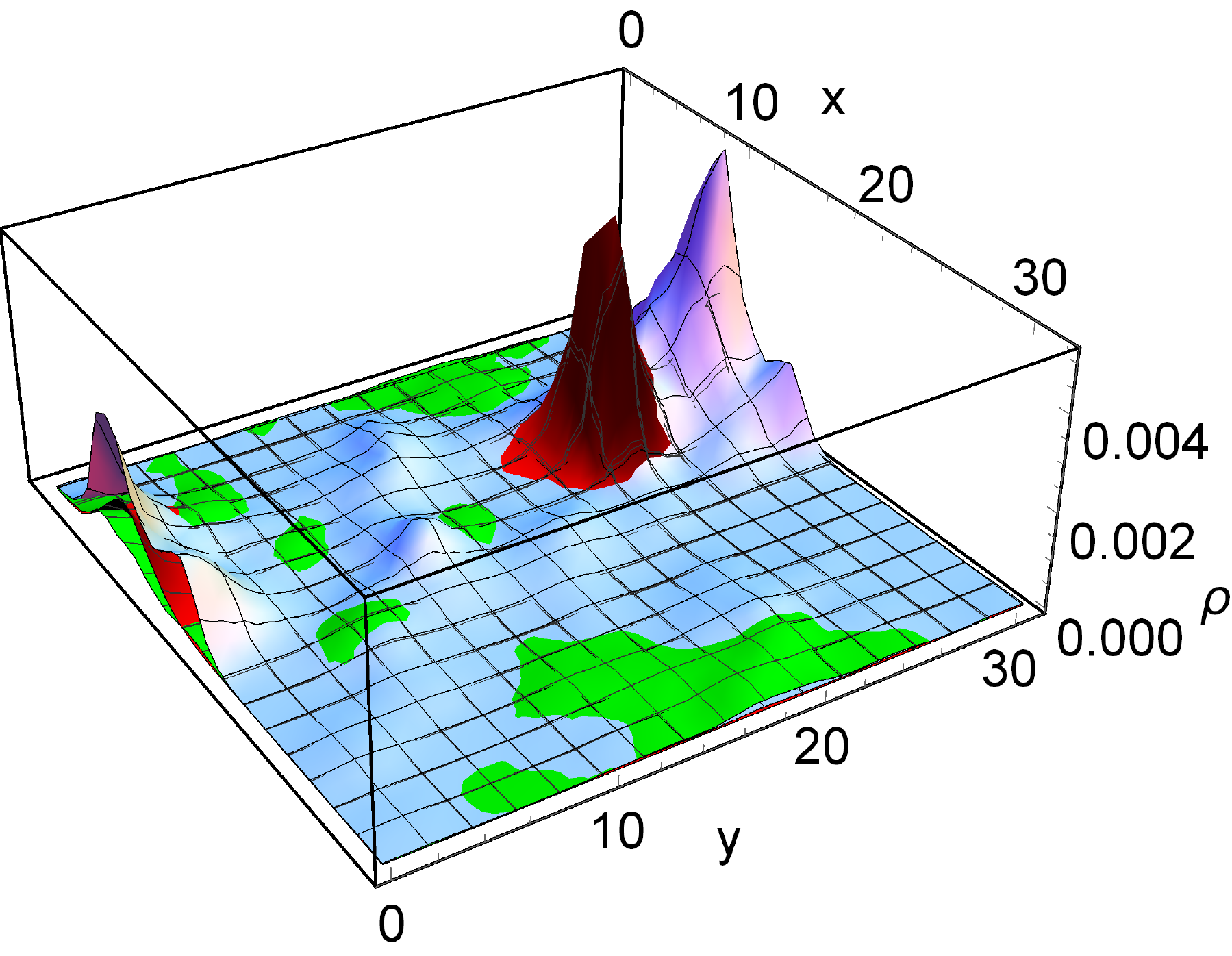}
\includegraphics[width=6cm]{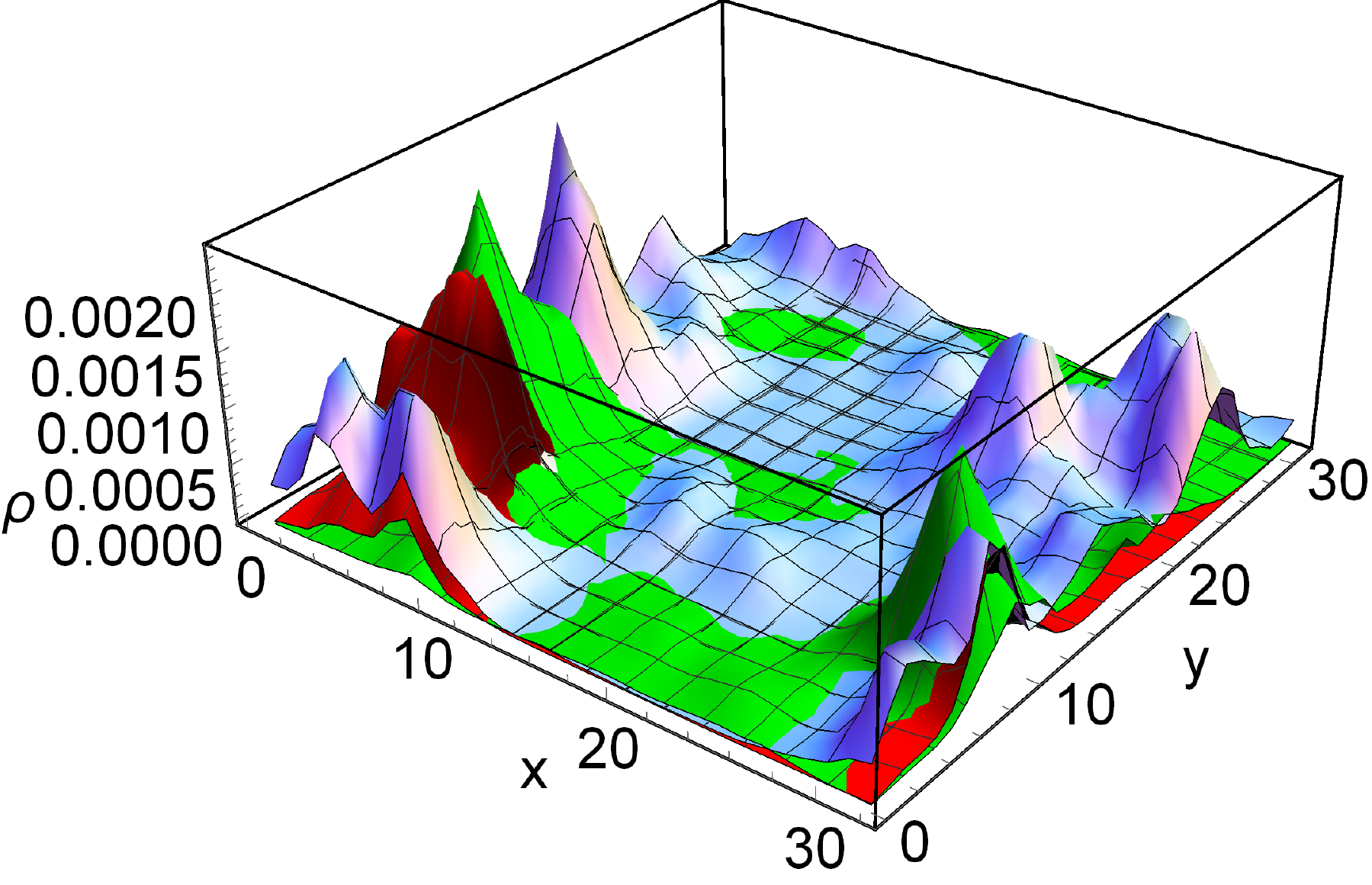}
\caption{Density $\rho(x,y)$ of delocalized zero modes for 
$\phi = \pi $ (red), $\pi /3$ (blue) and $-\pi /3$ (green) 
for $T/T_c=1, 1.08, 1.1$ at a fixed value of the coordinate 
$z$ at the maxima of the profile. The Euclidean time coordinate is 
summed over.  }
\label{fig:zmodedeloc}
\end{center}
\end{figure}

\begin{figure}[]
\begin{center}
\includegraphics[width=6cm]{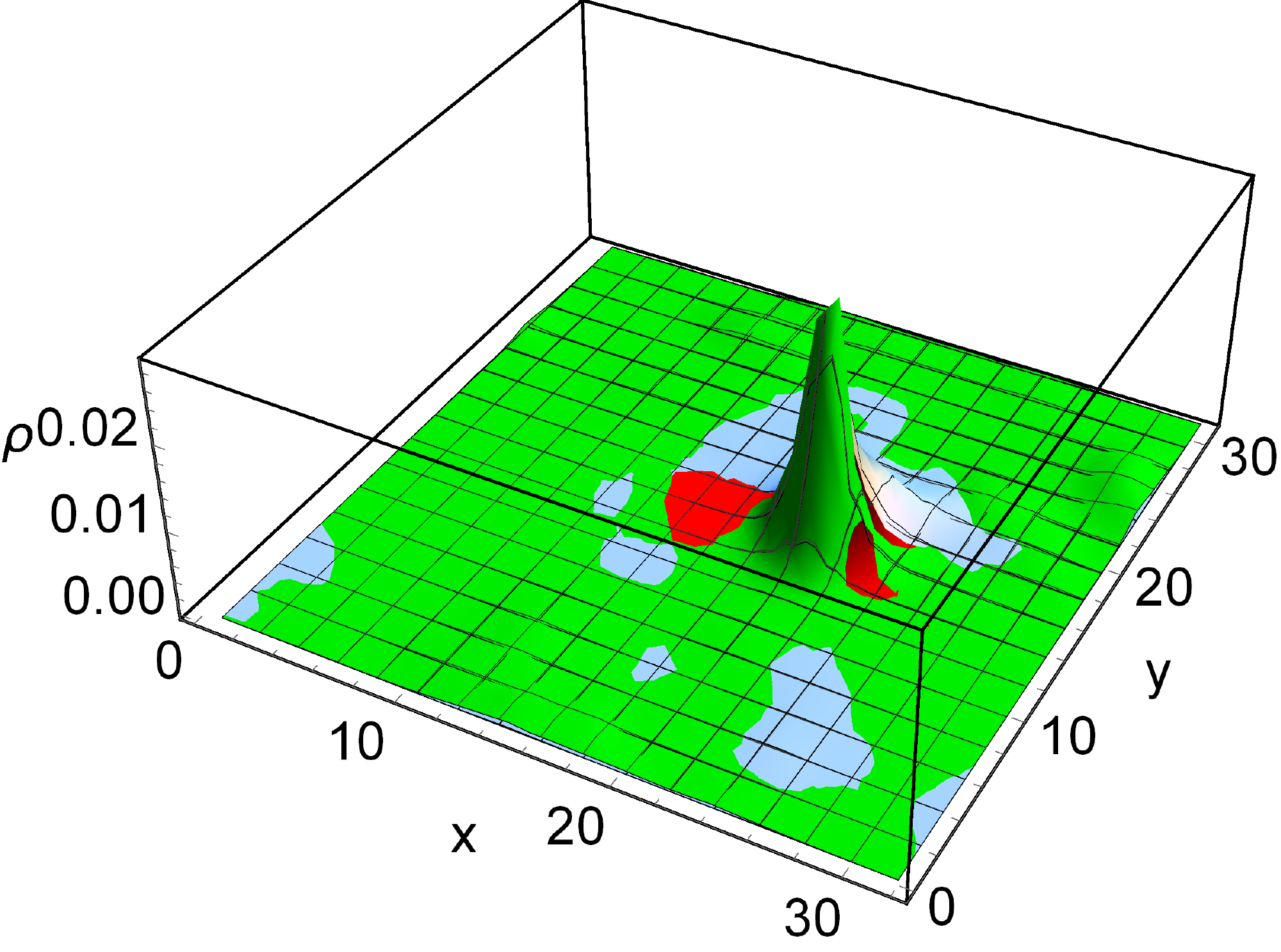}
\includegraphics[width=6cm]{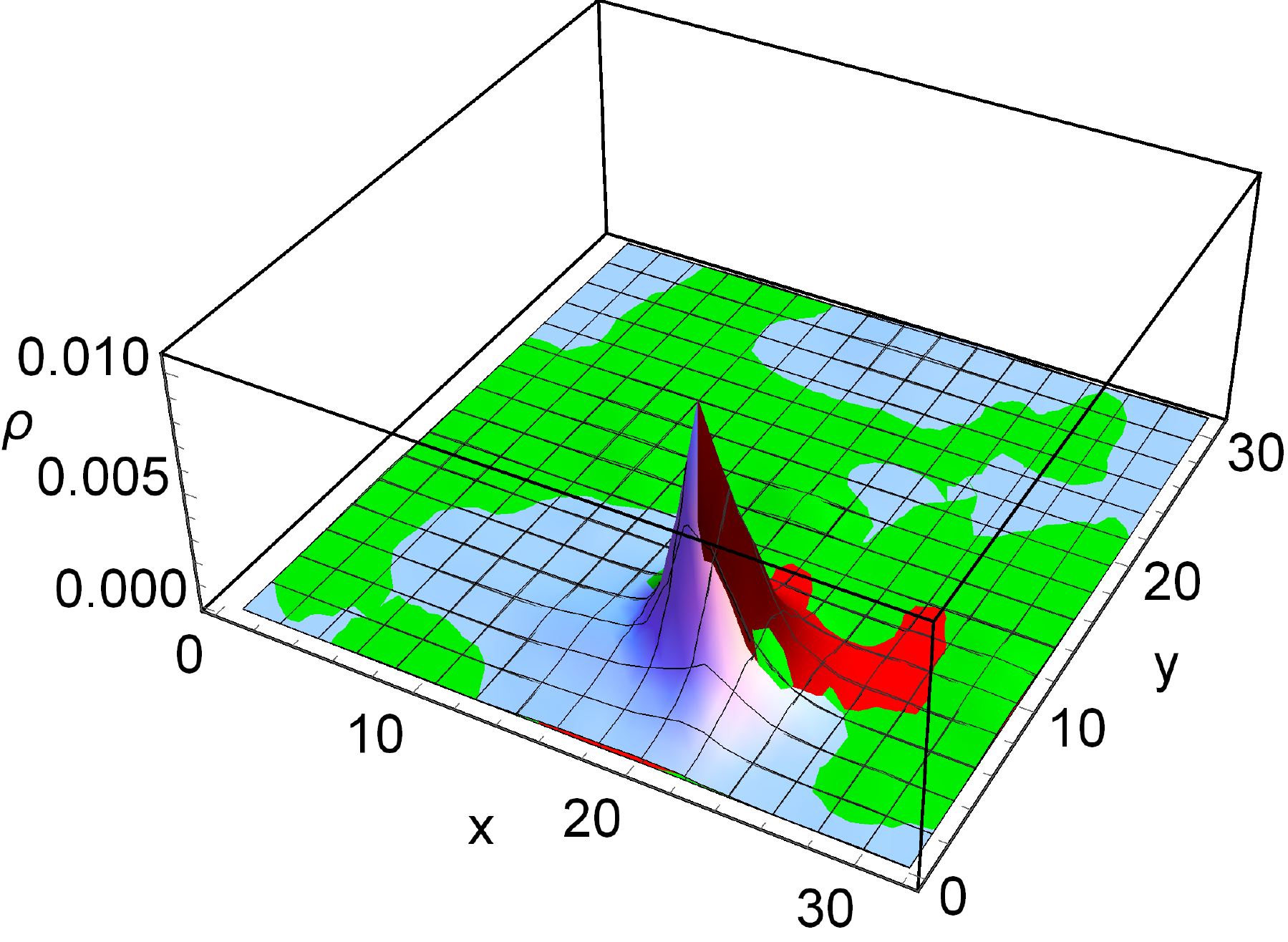}
\includegraphics[width=6cm]{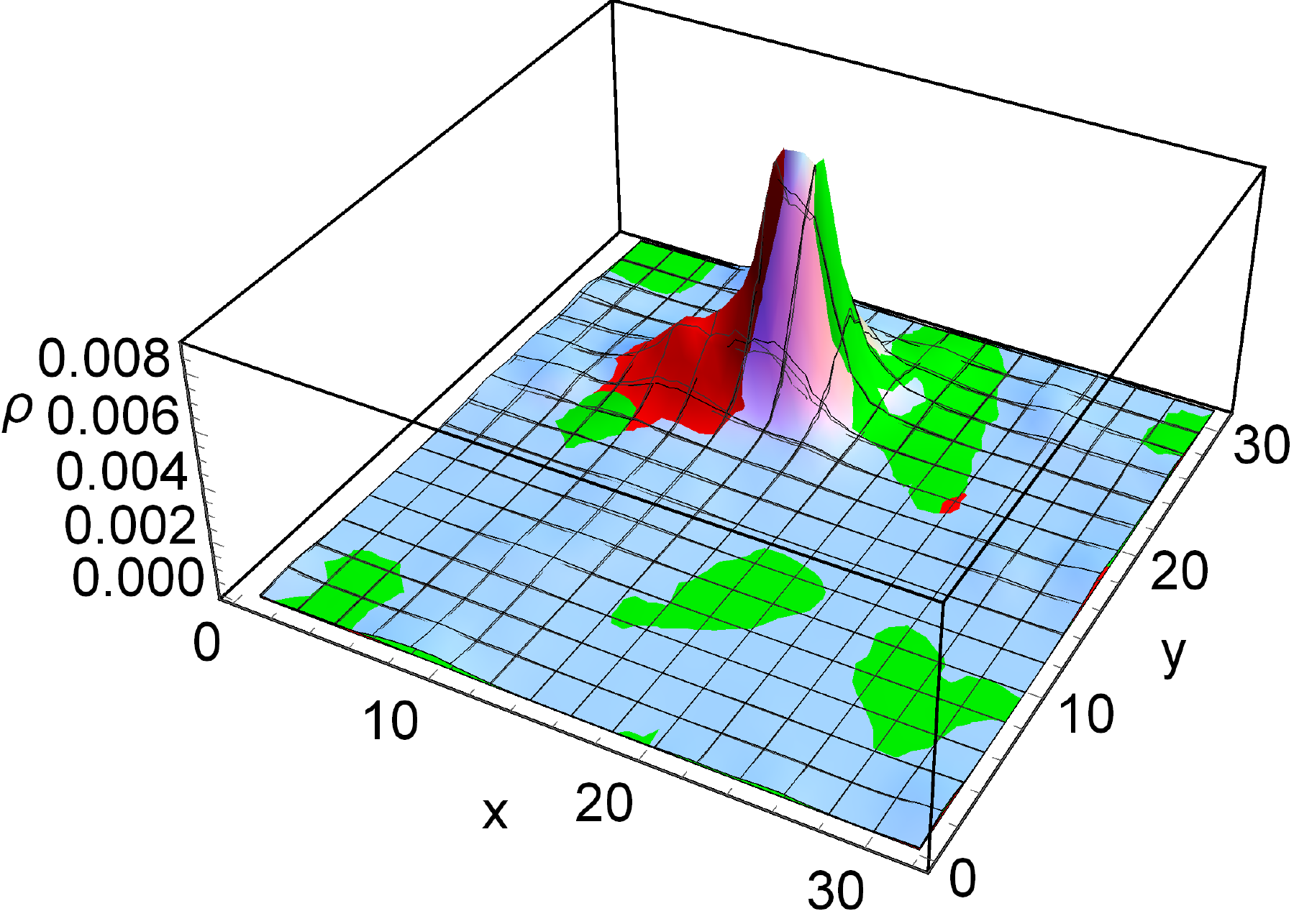}
\caption{Density $\rho(x,y)$ of localized zero modes for 
$\phi = \pi $ (red), $\pi /3$ (blue) and $-\pi /3$ (green) 
at $T/T_c=1, 1.08, 1.1$, and at a fixed value of the coordinate 
$z$ at the maxima of the profiles. The Euclidean time coordinate is 
summed over.}
\label{fig:zmodeloc}
\end{center}
\end{figure}

To answer this question, we look for the zero modes (of the valence  
overlap operator) for three different temporal periodicity phases 
$\phi=\pi,\pm \frac{\pi}{3}$. We study them on QCD ensembles at 3 
different temperatures between $T_c$-$1.1~T_c$. For all the 
statistically independent configurations considered, we have observed 
that both types of behavior mentioned above, are in fact present. Examples 
of such regimes are presented in Figures ~\ref{fig:zmodedeloc} and
 \ref{fig:zmodeloc} respectively. In Figure ~\ref {fig:zmodedeloc}, 
the peak positions of the quark zero-mode density changes for different 
temporal periodicity phases $\phi=\pi, \frac{\pi}{3}, -\frac{\pi}{3}$ 
providing us the first evidence that dyons are in fact observable as 
separate independent entities.  We observe similar behavior in a 
significantly large number of configurations at three different 
temperatures $T<1.2~T_c$, observing that the probability to find 
such configurations with distinct shift of zero modes corresponding 
to well separated dyons, reduces with temperature.

A contrasting picture of the \emph{topological cluster}
is shown in Figure ~\ref{fig:zmodeloc}. Here peaks of the 
zero-mode profiles corresponding to different fermion 
periodicity phases $\phi$ do \emph{not} shift. This may be 
due to two reasons:
\begin{enumerate}
\item[i)]
Firstly, the topological objects are still $L$ and $M$-dyons, 
but just located very close to each other. We remind that 
classically their mutual deformations keeps the total action 
constant, so any interaction between them should be of quantum 
nature.
\item [ii)]
Secondly, it can be that the local value of the Polyakov loop is 
trivial $\langle P \rangle \approx 1$ and thus the whole dyon
construction collapses back to instantons.
\end{enumerate}

We will unambiguously identify the exact nature of these zero modes 
in Sec.~\ref{sec:semiclassics}.  The next obvious question to address 
is the density of different types of dyons, and its dependence on 
temperature. Comparing ensembles at different temperatures, 
we observe that at the highest temperature $1.2~T_c$, 
the occurrence of fermion zero-modes decreases substantially. Moreover 
we rarely observe well separated zero mode profiles as we change the 
quark temporal-periodicity phase $\phi$. This may be due to the fact 
that at higher temperatures, the holonomy is trivial resulting 
in the average value of Polyakov loop to be close to unity. One of its 
eigenvalues $\nu_m=2\pi$, whereas two others are identically zero due to 
which only one type of dyon survive at high temperatures in statistically 
large number of configurations. We further elaborate on this case 
again in Sec.~\ref{sec:semiclassics}.

\subsection{Properties of the near-zero modes}

So far we have analyzed the most clean case of \emph{zero modes} 
in QCD configurations with $Q=\pm 1$. Now we proceed to
study the \emph{near-zero} modes in the same configurations. 
The spectral density of near-zero modes of the Dirac operator is an 
important observable since it is directly related to the chiral condensate 
through the Banks-Casher relation. Depletion of their density near-zero 
eigenvalues thus means disappearance of this condensate and the restoration 
of the $SU_A(N_f)$ chiral symmetry. A separate important topic is the fate 
of anomalous $U_A(1)$ chiral symmetry in this temperature range, also 
influenced by a certain functional dependence of the near-zero eigenvalue 
spectrum.  The very appearance of Dirac eigenvalues close to zero is due to 
\emph{collectivization} of a large number of topological objects, scaling 
with the lattice volume $\sim O(V_4)$. This phenomenon is related to the 
density of such objects, but in a subtle indirect way, since topological 
objects can make neutral clusters with zero overall topological charge. 
Since near-zero modes arise as a result of interactions of the underlying 
topological objects, these thus contain crucial information on their nature 
and the density. 

\begin{figure}[]
\begin{center}
\includegraphics[width=6cm]{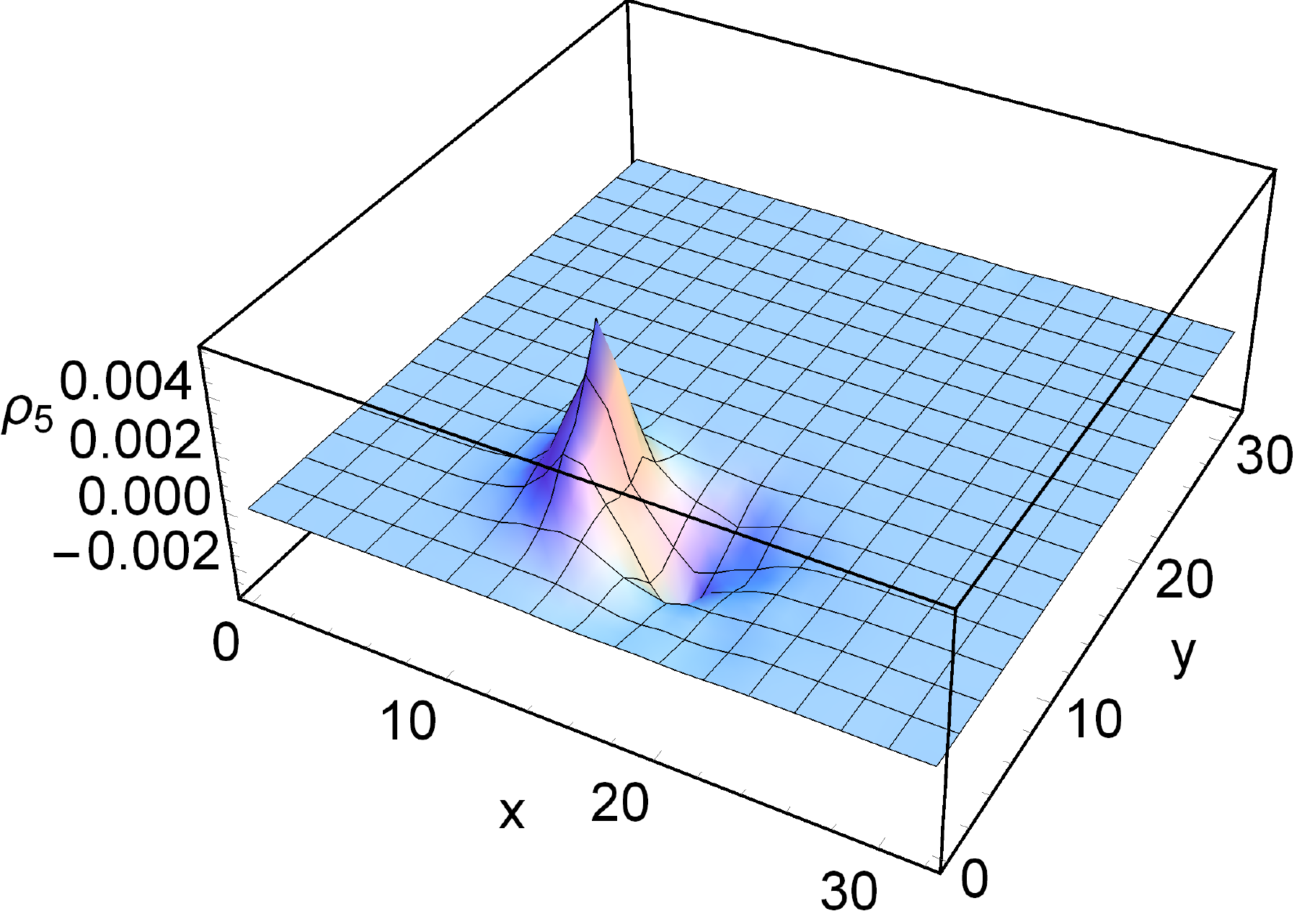}
\includegraphics[width=6cm]{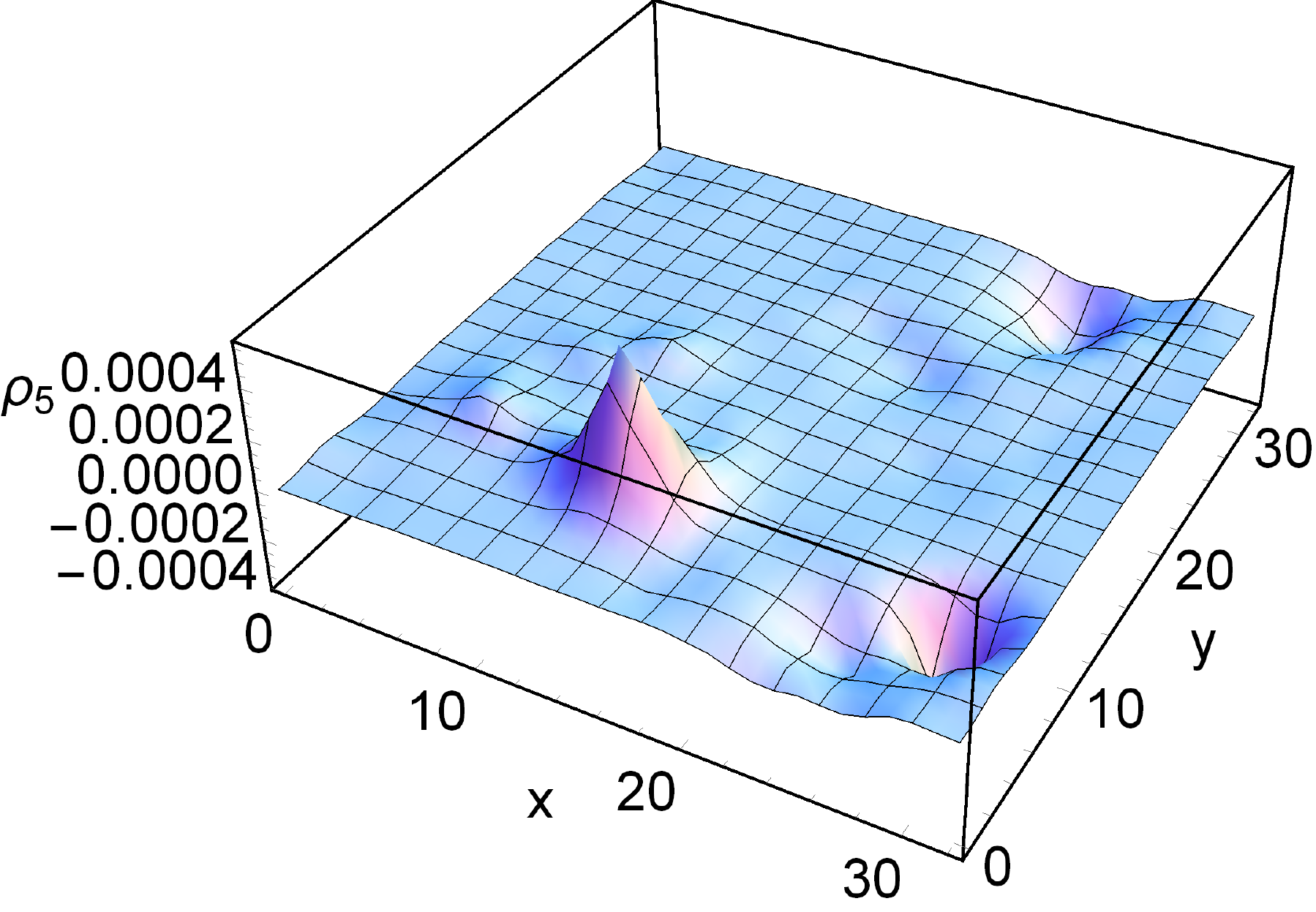}
\caption{The chiral density $\rho _5(x,y)$ profile for the first 
near-zero mode at $T=1.1~T_c$ for periodicity phase $\phi=\pi$ 
(top panel) and $\phi=\pi/3$ (bottom panel) respectively. The 
$z$ and $\tau$ coordinates are fixed to the maxima of the peaks. }
\label{fig:nzeromode108tc}
\end{center}
\end{figure}

The first qualitative question we address here, is whether
the density of different species of dyons are the same or not, 
as a function of temperature. For this we measure the chiral 
density profile $\rho_5(x,y)$ of the first near-zero mode as 
a function of two spatial coordinates, with the $z$- coordinate 
fixed and the temporal coordinate summed over. It is very striking 
that the densities of $L$ and $M$-type dyons, observed via peaks in 
the chiral density profiles of the near-zero modes, are indeed very 
different!  
As an example, let us show the near-zero mode $\rho_5$-profiles 
for the configuration at $1.1~T_c$. 
We plot the chiral density profiles for two different periodicity 
phases $\phi=\pi, \pi/3$ in the upper and the lower panels of 
Figure~\ref{fig:nzeromode108tc}, respectively. In the case  
$\phi=\pi$ we observe only one closely located dyon and an 
anti-dyon pair, while for the case when $\phi=\pi/3$, we 
observe more than one dyon and anti-dyon pairs, both close 
as well as widely separated.

To explain these results, let us recall that the finite temperature 
QCD ensembles which are studied here were generated using the 
usual anti-periodic boundary conditions for the quark fields 
along the temporal direction. Hence the presence of weakly-interacting 
widely separated $L$-dyon pairs are already suppressed due to the fact 
that in such cases the quark determinant is close to zero. 
However widely separated $M$-dyon pairs (which do not correspond 
to anti-periodic boundary conditions along the temporal direction) can 
still exist with a small but finite probability, in all such 
ensembles. When we probe such ensembles with valence overlap 
fermions with quark periodicity phase $\pi$ (corresponding to $L$-dyons) 
then its near-zero modes arises only due to closely separated pairs 
of $L$-dyons and anti-dyons. Changing the valence quark periodicity 
phase to $\phi=\pi/3$, its near-zero modes measures the $M$-dyon 
pairs which may be either close or widely-separated, all of these 
cases contribute to the near-zero spectrum. To summarize, the near-zero 
modes of the Dirac operator are excellent probes of the identity and 
the interactions between dyons. It allows us to distinguish between 
$L$ and $M$-dyon species unambiguously and also to estimate the 
interactions between them, which we show through this example.

\subsection{ Spectral density of the Dirac eigenvalues 
with different temporal periodicity phases}

In the preceding section we have shown that the $L$-dyons are paired, and 
experience relatively stronger interaction,  compared to the $M$-dyons. Now 
we proceed from qualitative examples to a full statistical sampling of all 
near-zero modes we have, making this comparison more quantitative.

\begin{figure}[]
\begin{center}
\includegraphics[width=8cm]{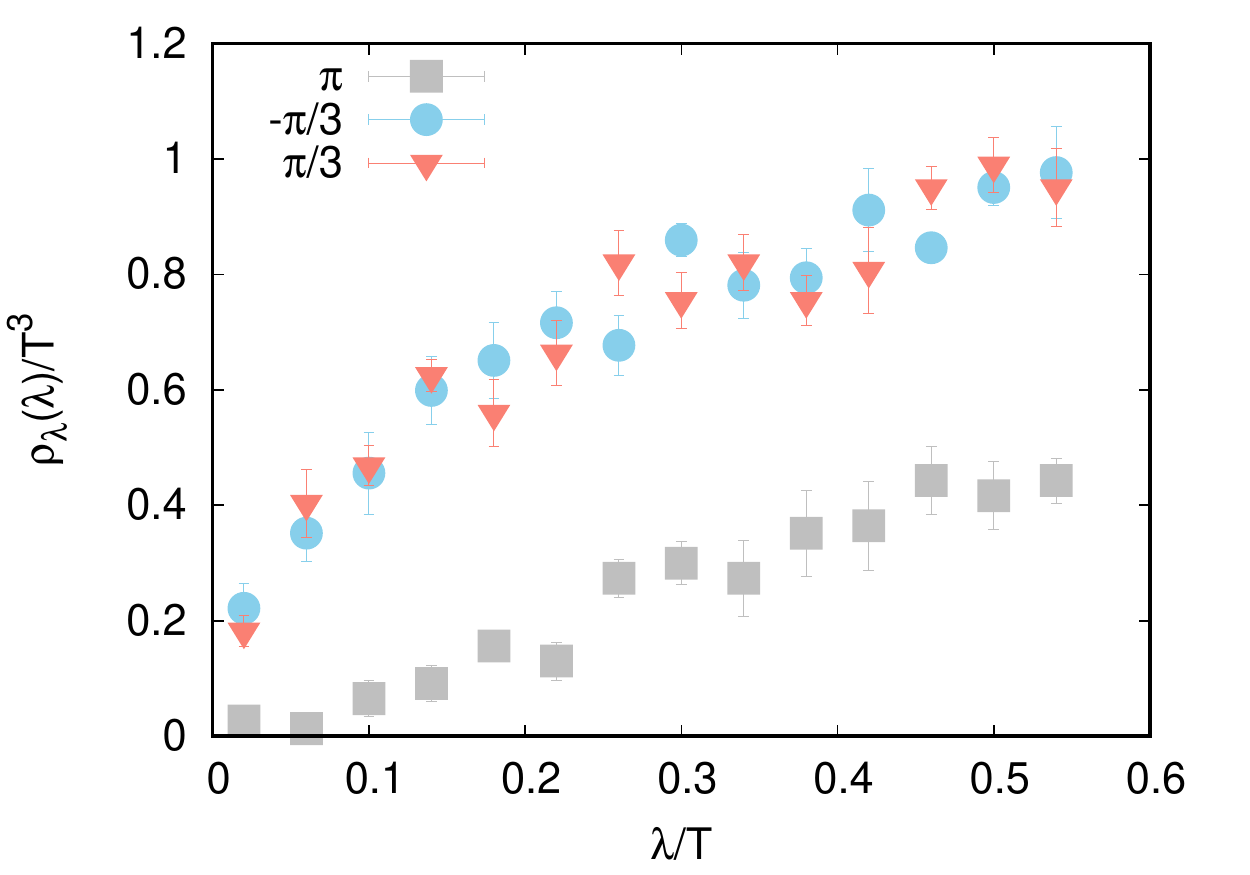}
\includegraphics[width=8cm]{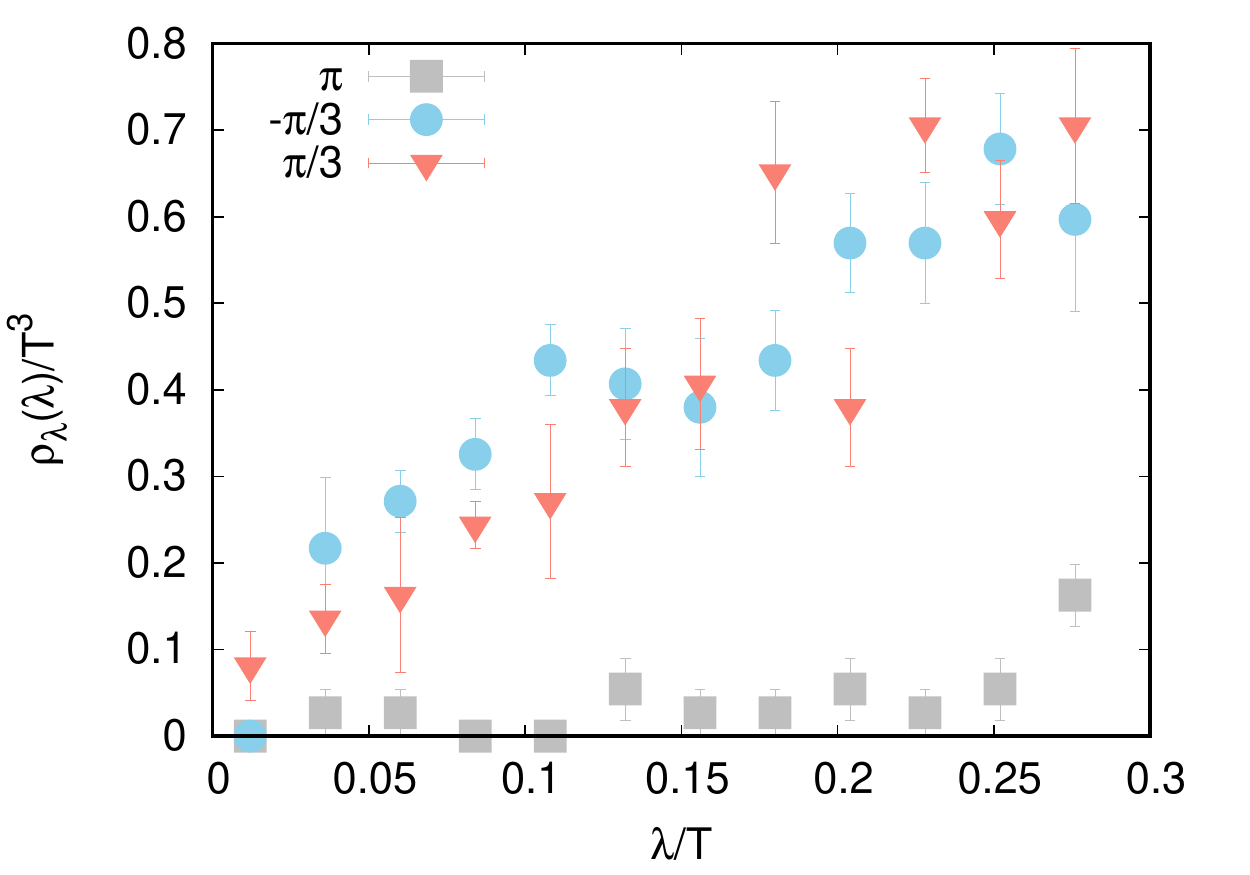}
\caption{The spectral density of the (valence overlap Dirac) eigenvalues 
$\rho(\lambda)/T^3$  at $T=1.1~T_c$ (top panel) and $T=1.2~T_c$ (bottom panel) 
for different quark temporal-periodicity phases $\phi=\pi, \pm\pi/3$, respectively.}
\label{fig:nearzerob1771}
\end{center}
\end{figure}

The spectral density of the Dirac eigenvalues, of the valence 
overlap operator for  QCD gauge configuration at $1.1~T_c$ is 
shown in the top panel of Figure~\ref{fig:nearzerob1771}. Three 
sets of points correspond to different quark periodicity phases. 
Clearly one observes from it that an ensemble of $L$-dyons, 
generating near-zero eigenvalues shown in square symbols, 
is less dense than that by $M$-dyons (circles and triangles). 
Furthermore, the corresponding quark condensates (corresponding 
to the spectral density of, near-zero eigenvalues which extrapolate 
to zero in the infinite volume limit, $\rho_\lambda(\lambda\rightarrow 0)$) 
vanishes for $L$-dyons and remains finite for $M$-dyons. So, 
in the real world, in which quarks are fermions and the physical 
quark condensate is generated due to the $L$-dyons, this temperature 
is clearly in the chiral symmetry restored phase. And yet, at the 
same temperature the ensemble of $M$-dyons is dense enough to still 
be \emph{below} its chiral restoration temperature! The same statement 
holds for our highest temperature $T=1.2~T_c$ as well, as evident from 
the eigenvalue spectrum shown in bottom panel of Figure~\ref{fig:nearzerob1771}.

In order to appreciate how different the ensembles of $L$ and $M$ dyons are, 
we note that the spectral density is shown up to rather large eigenvalues, 
$\lambda\sim 60$-$80$ MeV. In other words, the number of low-lying eigenstates 
we calculated is $\sim 30$ per gauge configuration, a factor $3$-$5$ larger 
than the mean number of dyons of all kinds in them. This means that the 
zero-mode zone(ZMZ) generated by such topological solitons is responsible 
only for a fraction of the spectral density corresponding to roughly 
$\sim\lambda/T < 0.2$. Moreover, the difference between the spectral 
densities of $L$ and $M$-dyons is evidently quite large. 
Here the eigenvalue density is calculated after performing
statistical averaging of more than $30$ topologically 
independent configurations in order to ensure that all 
allowed topological charge sectors are ergodically 
sampled. Hence it is a more robust observable insensitive to 
topological freezing and gauge field fluctuations. 

To summarize this section, since at higher temperatures only 
closely located $L$ dyon-antidyon pairs exist,there is a 
significant depletion of the corresponding near-zero quark 
modes. However since the $M$-dyon pairs of all separations 
survive, the corresponding fermion near-zero density does not 
fall with temperature as drastically as that of the $L$-dyons. 
We will measure the distribution of the separation lengths 
between $L$ and $M$-dyon pairs in a subsequent section.

\subsection{How robust are the small eigenmodes to UV lattice artifacts}

In this section we provide some evidences that the cut-off effects 
in our studies are under control and our conclusions in the previous 
sections are physically robust. To this end, we study the zero and 
near-zero modes of the valence overlap Dirac operator for different 
quark-periodicity phases at $\sim T_c$ and $1.08~T_c$ on the gauge 
configurations used previously, but now with the highest modes 
of the gauge fields (momenta $\sim 1/a$) removed.

\begin{figure}[]
\begin{center}
\includegraphics[width=6cm]{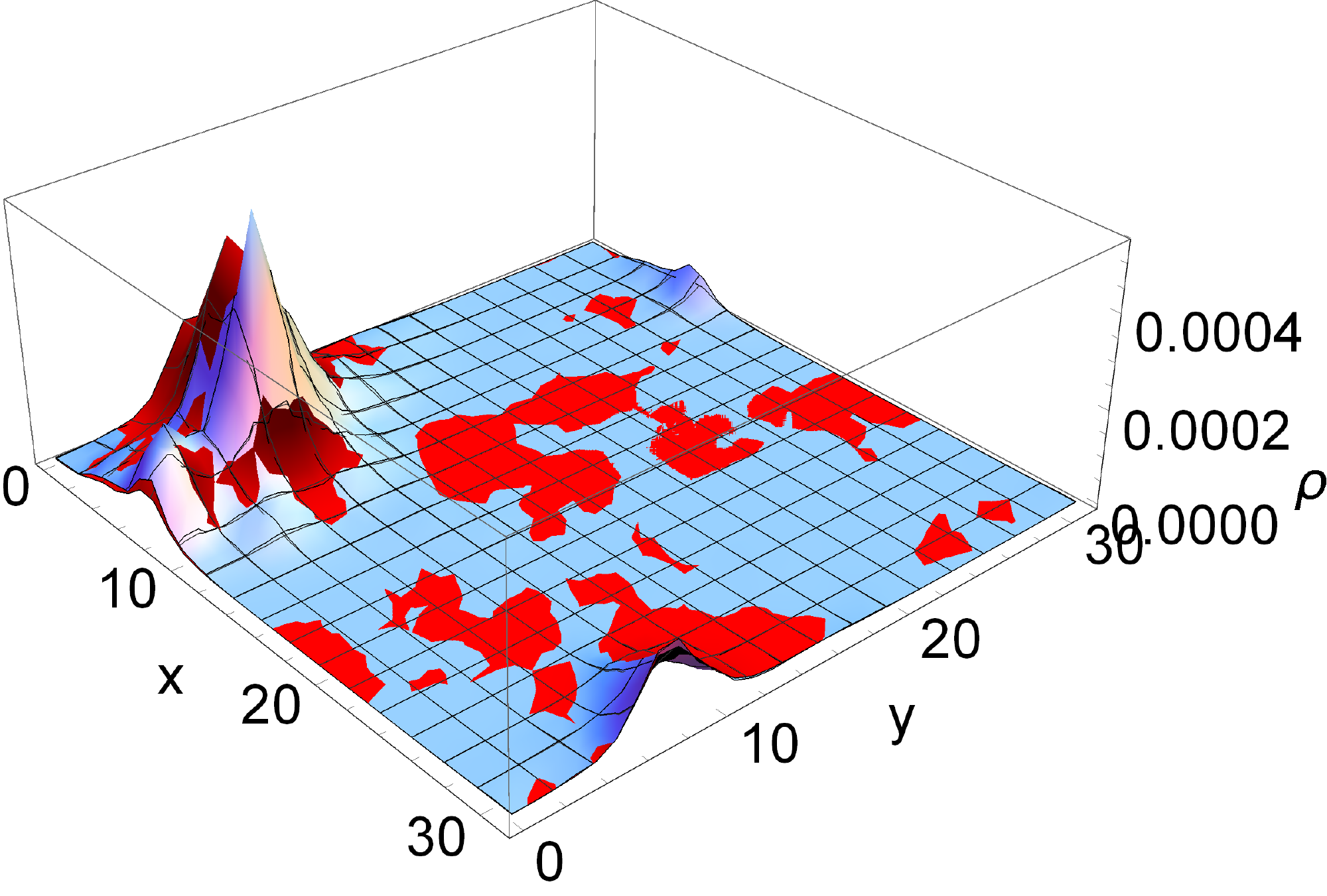}
\includegraphics[width=6cm]{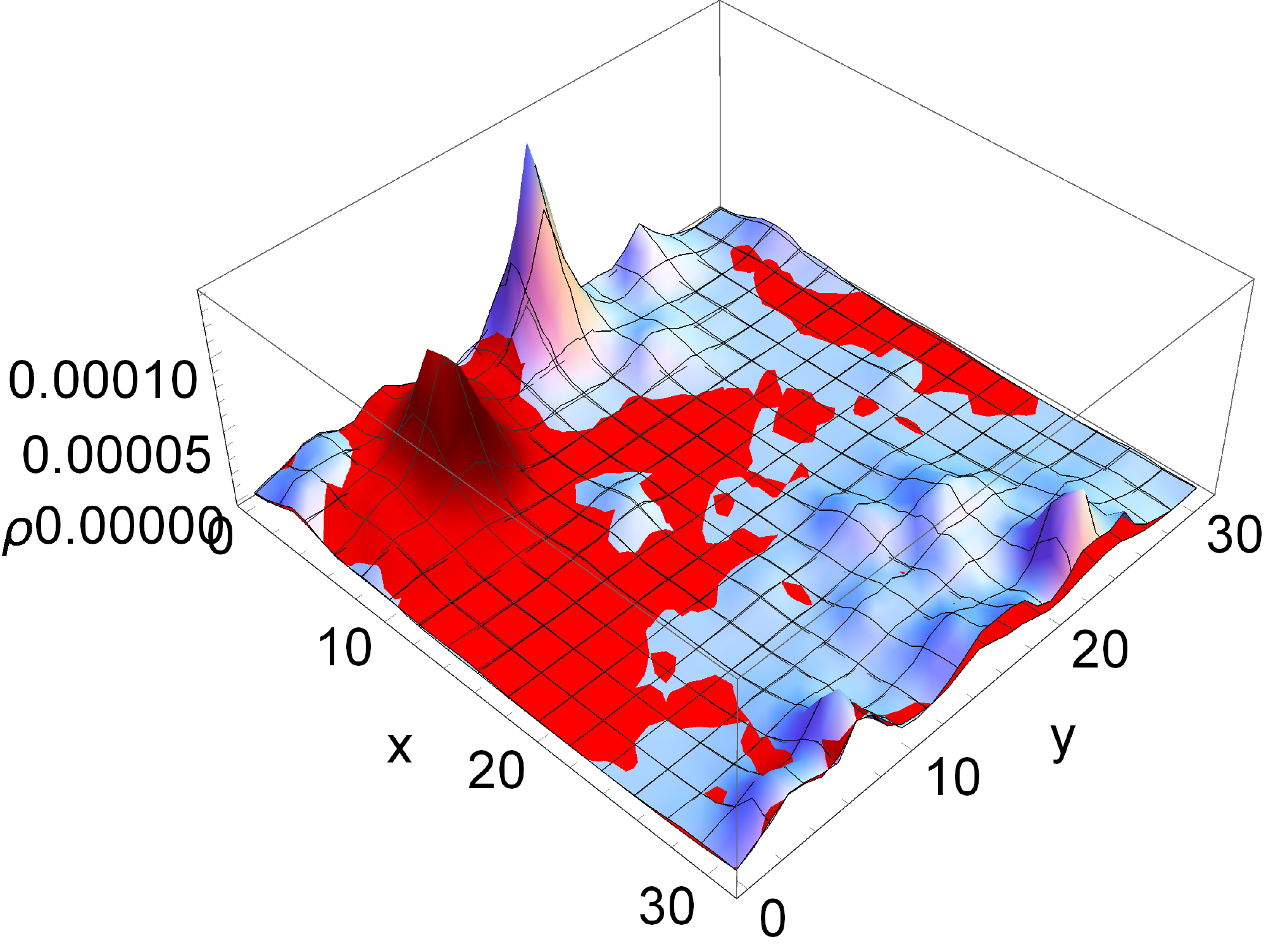}
\includegraphics[width=6cm]{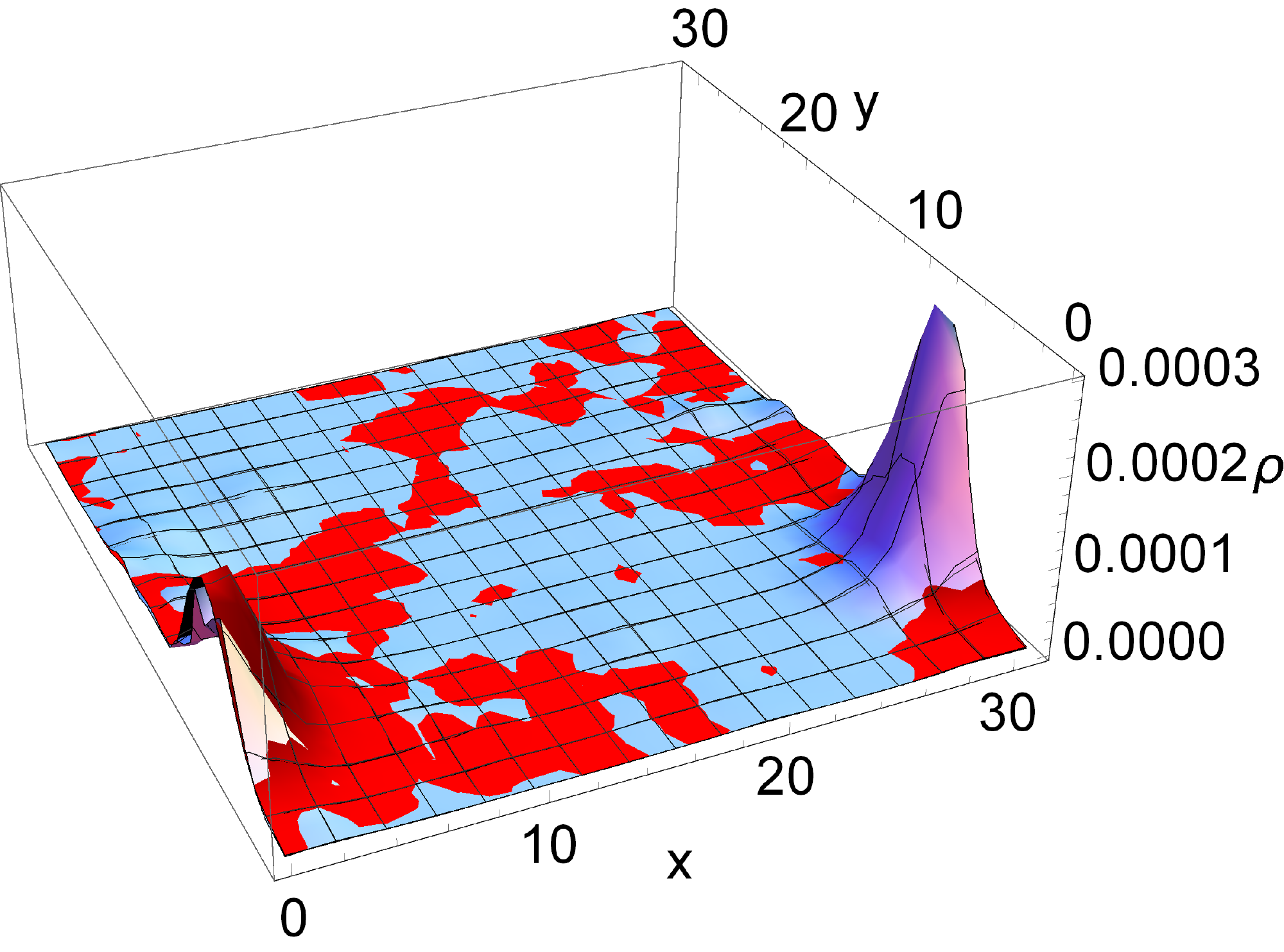}
\caption{The valence overlap quark zero-mode densities measured 
on a two-step HYP smeared (red) and the original un-smeared (blue) 
gauge configuration for three different quark periodicity phases 
$\phi=\pi$ (top panel), $\phi=\pi/3$ (middle panel) and $\phi=-\pi/3$ 
(lower panel), shows a remarkably good agreement. This is a configuration 
at $1.1~T_c$ with a net topological charge $Q=1$. }
\label{fig:zeromodes-smeared-b1771}
\end{center}
\end{figure}

To implement this we performed two steps of HYP smearing~\cite{Hasenfratz:2001hp} 
on the above mentioned gauge configuration in order to smoothen the 
ultra-violet fluctuations, which exist on the length scale of one 
unit of lattice spacing. If the zero modes we observed earlier were 
due to gauge dislocations i.e., localized artifacts of typical extent 
of about a lattice spacing, which are known to occur with a finite probability 
at coarser lattice spacings, then they should disappear after smearing. 
The topological objects like dyons and calorons are of typical sizes of 
several lattice spacings (for fine enough lattice spacings they can 
extend up to several lattice sites) and hence their topological properties 
should survive such smearing, in fact, any such continuous deformations.
Indeed the ultra-violet modes of the gauge fields seems not to affect the 
shape and location of the zero-modes, as is evident from the comparison of 
the zero-mode profiles at $1.1~T_c$ shown in Figure~\ref{fig:zeromodes-smeared-b1771}, 
before and after minimal smearing and for three different quark temporal-
periodicity phases. Furthermore the height, width and the number of zero-modes 
of the fermions of each periodicity phase $\phi$, remain unchanged after the 
ultra-violet smearing, which indicates that these are not artifacts due to 
finite lattice spacing. These conclusions survive even at temperatures close 
to $T_c$.  The near-zero modes too remain unaffected from the smearing of 
the ultra-violet fluctuations leading us to conclude that the lowest modes 
of the Dirac operator we use are indeed \emph{insensitive} to lattice cut-off 
effects. 

\subsection{Multi-parameter fits of the topological clusters}
\label{sec:hol}
To establish connections between our lattice results for the 
quark zero-mode density and the corresponding analytic expression
in Eq.~\ref{eqn:zmode1}, we calculate the latter numerically 
by constraining the locations and widths of the dyons from the 
lattice data. There are several input parameters needed to initiate 
the fit routine. Among these parameters, the important 
ones are the holonomies i.e., the Polyakov loop phases $\mu_i,i=1,2,3$, 
since these define the actions (and thus topological charges) as well 
as physical sizes of the dyons. The details of the fitting procedure 
are rather technical and are mentioned in Appendix A for the interested 
readers. Here we present here only the results of direct physical interest.

\begin{figure}[]
\begin{center}
\includegraphics[width=8cm]{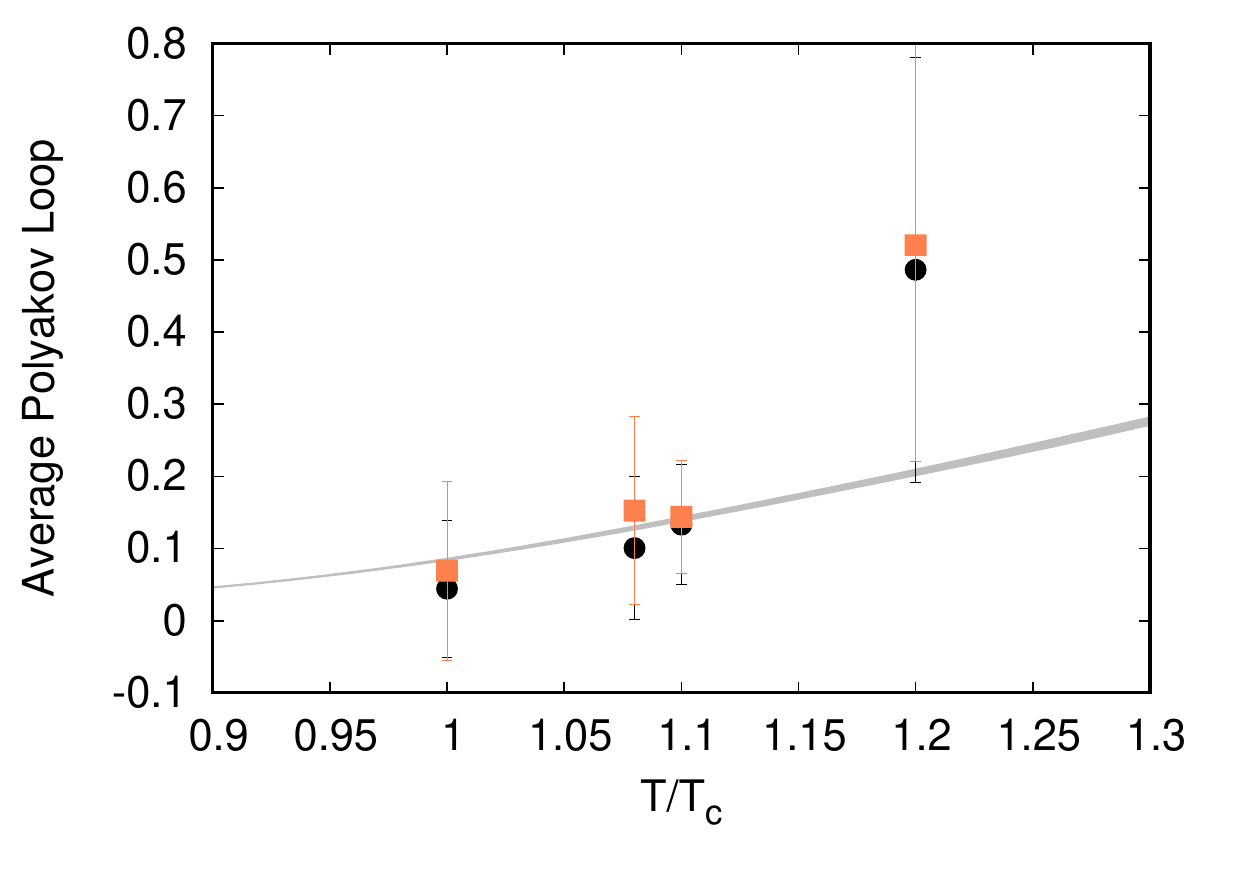}
\caption{The final average value of Polyakov loop obtained after fitting the 
lattice zero-mode profiles to the semi-classical results at $\phi = \pi$. The 
two different data points in black and orange correspond to two different initial 
choices for the average Polyakov loop $\langle P \rangle=0$ and $\langle P \rangle=1/3$, 
respectively to initiate the fitting procedure. The gray band is the continuum 
extrapolated value of Polyakov loop in QCD known from independent lattice 
studies~\cite{Bazavov:2016uvm,Bazavov:2018wmo}.}
\label{fig:AvPolyakovLoop}
\end{center}
\end{figure}

Among the obvious questions to raise here are whether the values
of the holonomies obtained from the fits do show a tendency to vanish 
at $T_c$ (the lowest of our temperature range), and how well the values 
of the Polyakov loop we get agree with what is known from its continuum estimate 
in QCD obtained from independent lattice studies. Our results for final values 
of $\langle P \rangle$ obtained as a result of the fits are shown in 
Figure~\ref{fig:AvPolyakovLoop}. Agreement between two set of points demonstrate 
convergence of the fits, starting from very different initial trial values, 
$\langle P \rangle=0$ (circles) and $\langle P \rangle=1/3$ (squares). 
Furthermore, it displays the expected trend of having a vanishingly 
small value at $T_c$, growing to $\langle P \rangle\sim 1/2$ at our highest 
$T=1.2~T_c$, although with a large error as the number of configurations (used 
for the fit) with $|Q|=1$ at $T=1.2~T_c$ are quite limited. The final values at 
each temperature obtained from the fits, agrees quite well with the continuum 
estimates of the Polyakov loop obtained from independent lattice studies~\cite
{Bazavov:2016uvm,Bazavov:2018wmo}, shown as a gray band in the same Figure.

\begin{figure}[]
\begin{center}
\includegraphics[width=8cm]{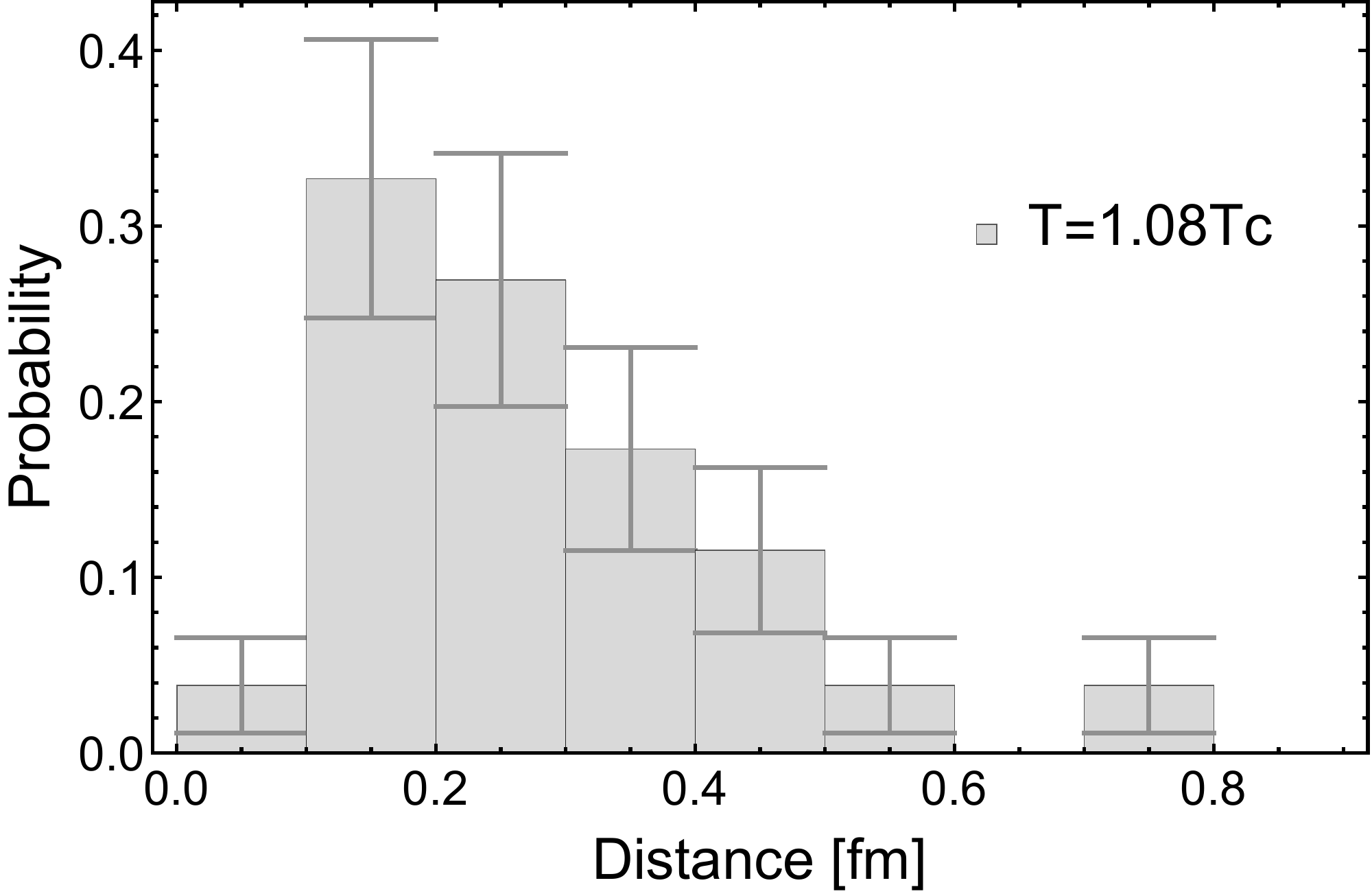}
\includegraphics[width=8cm]{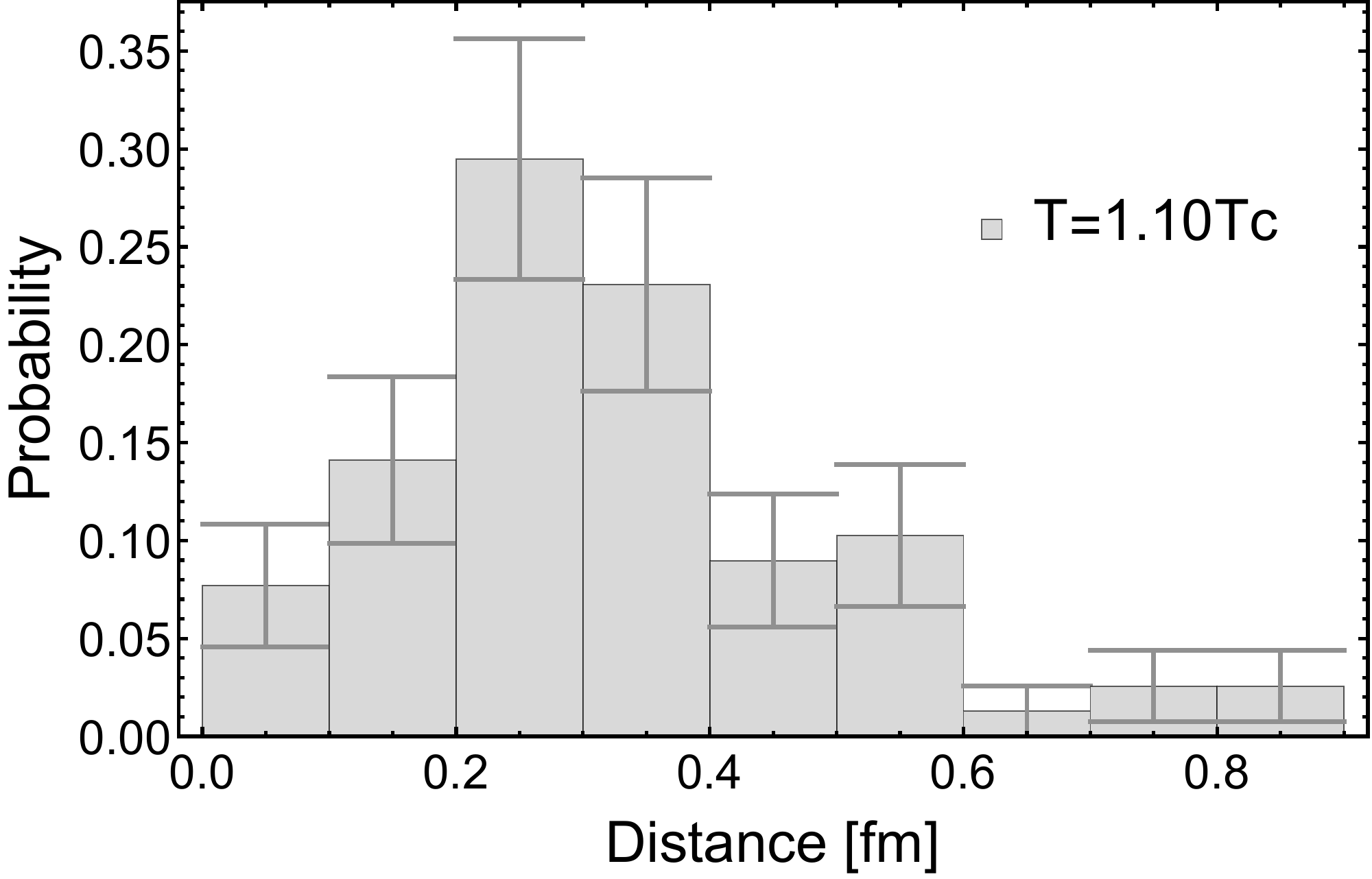}
\caption{The histogram of the separation distances between a $L$ and a $M$-dyon 
at $T/T_c=1.08$ (upper panel) and $T/T_c=1.1$ (lower panel).}
\label{fig:DistanceLtoMHist}
\end{center}
\end{figure}

The other characteristic feature of the topological clusters obtained 
as a result of the fits, are the  distribution of the dyon locations. With 
a triad of $L, M_1, M_2$ dyons identified, one can measure the distances between 
the centers of a $L$ and any one of the $M$-dyons. The histograms of the 
probability distribution of the distances between these two species of 
dyons is shown in Figure~\ref{fig:DistanceLtoMHist}. The most probable 
distance between a $L$ and $M$-dyon pair is between $0.2$-$0.3$ fm at 
$1.1~T_c$ whereas at $1.08~T_c$ it seems to be lower but within errors, 
are equally probable in the earlier range. Within the 
statistical errors of our data, we do not observe any significant temperature 
dependence of the topological cluster sizes. Thus with reasonable consistency 
the average separation between $L$ and $M$-dyons in the temperature range studied 
so far, is about $\sim 0.3$ fm.

\begin{figure}[]
\begin{center}
\includegraphics[width=8cm]{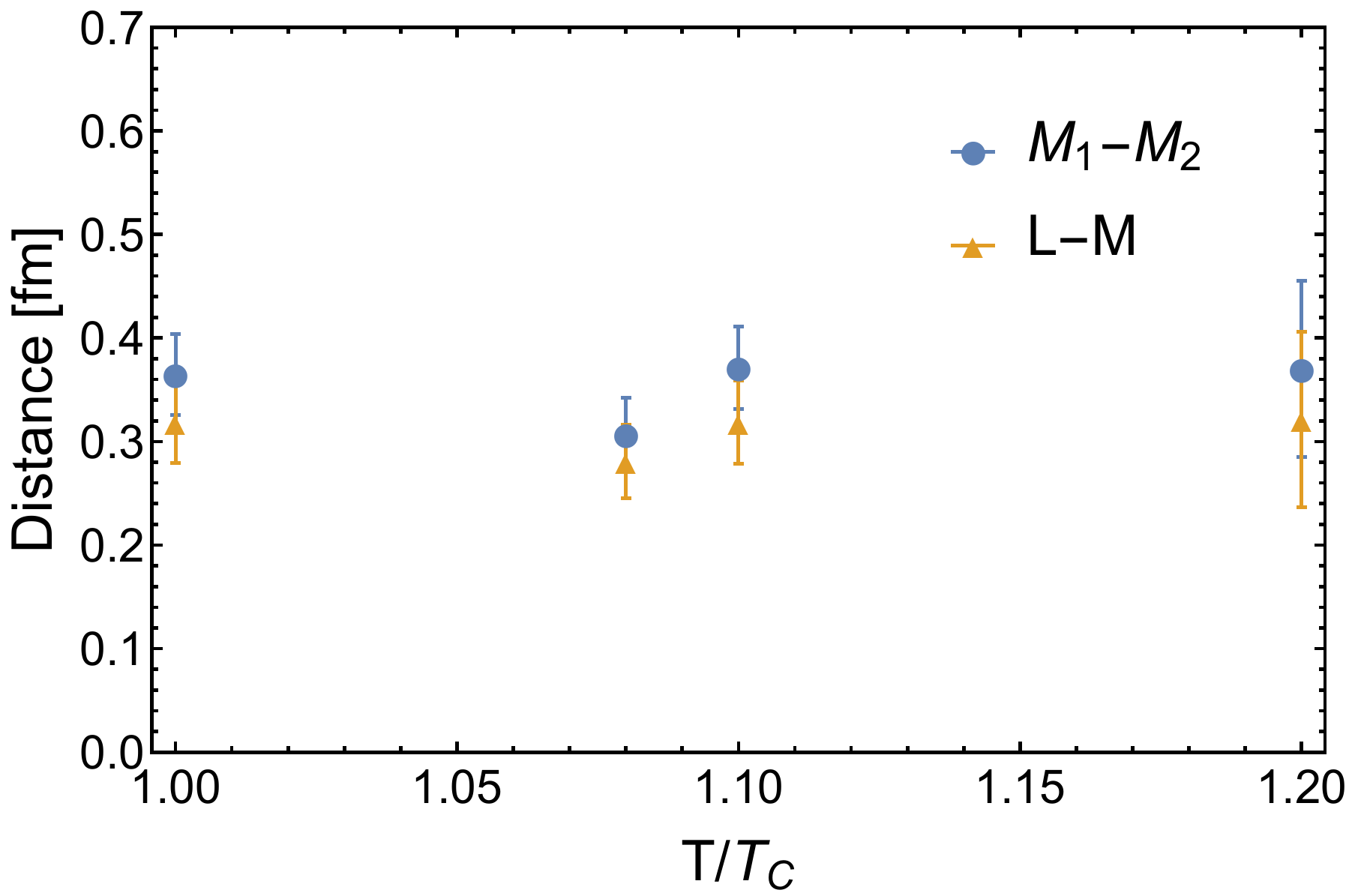}
\caption{Average distance in Fermi between centers of  $L$ and $M$ dyons 
(triangles) and $M_1$ and $M_2$ dyons (circles) for well-resolved topological 
clusters.}
\label{fig:DistanceLtoM}
\end{center}
\end{figure}

 We also compare the average distances between a $L$ and a $M$-dyon to that 
 between two $M$-dyons, results of which are shown in Figure~\ref
{fig:DistanceLtoM}.  At all the temperatures above $T_c$, the average 
distance between $M$-dyons seems to be always larger but at present the 
errors on the values are quite large to make very precise predictions. 
Since the relation between inter-dyon distance $r_{ML}$ and the caloron size
parameter is $\rho^2={ r_{ML} \over \pi T} $~\cite{Diakonov:2004jn},
the value, $r_{ML}\approx 0.3$ fm obtained from our study is in agreement 
with mean instanton sizes $\rho\sim 1/3$ fm of the Instanton Liquid Model 
of the QCD vacuum~\cite{Shuryak:1981fza}.

A few more additional comments regarding the fits are as follows:
\begin{enumerate}
\item [i)]
We remind that within a 13-dimensional parameter space of our fit ansatz, 
it is non-trivial to determine the functional dependence of the $\chi^2$ 
landscape. It could well be possible for the final set of parameters 
obtained after the $\chi^2$ minimization, to end up in an entirely different 
minima. Agreement between the two different fit results implies that, at 
least on average, we do not get different results for very different initial 
assumptions of parameters and the fitting procedure is reliable with good 
convergence property.

\item [ii)]
We usually did not consider the final fit values from those ensembles where 
the dyons of the same type are too close to each other e.g. in 
Figure~\ref{fig:DistanceLtoM}, as the fit results did not yield a 
low $\chi^2$.
\end{enumerate}

We conclude the section with a very optimistic observation that
within (admittedly large) statistical errors, our fitted values for 
the average value of the Polyakov loop are quite consistent with the 
average QCD data at all temperatures studied. Furthermore the fits to 
the distances between dyons quantify the notion of \emph{ionization} 
of instantons, to a typical size of a femtometer.

\subsection{Is the semi-classical theory of dyons accurate near $T_c$?}
\label{sec:semiclassics}

\begin{figure}[]
\begin{center}
\includegraphics[width=7cm]{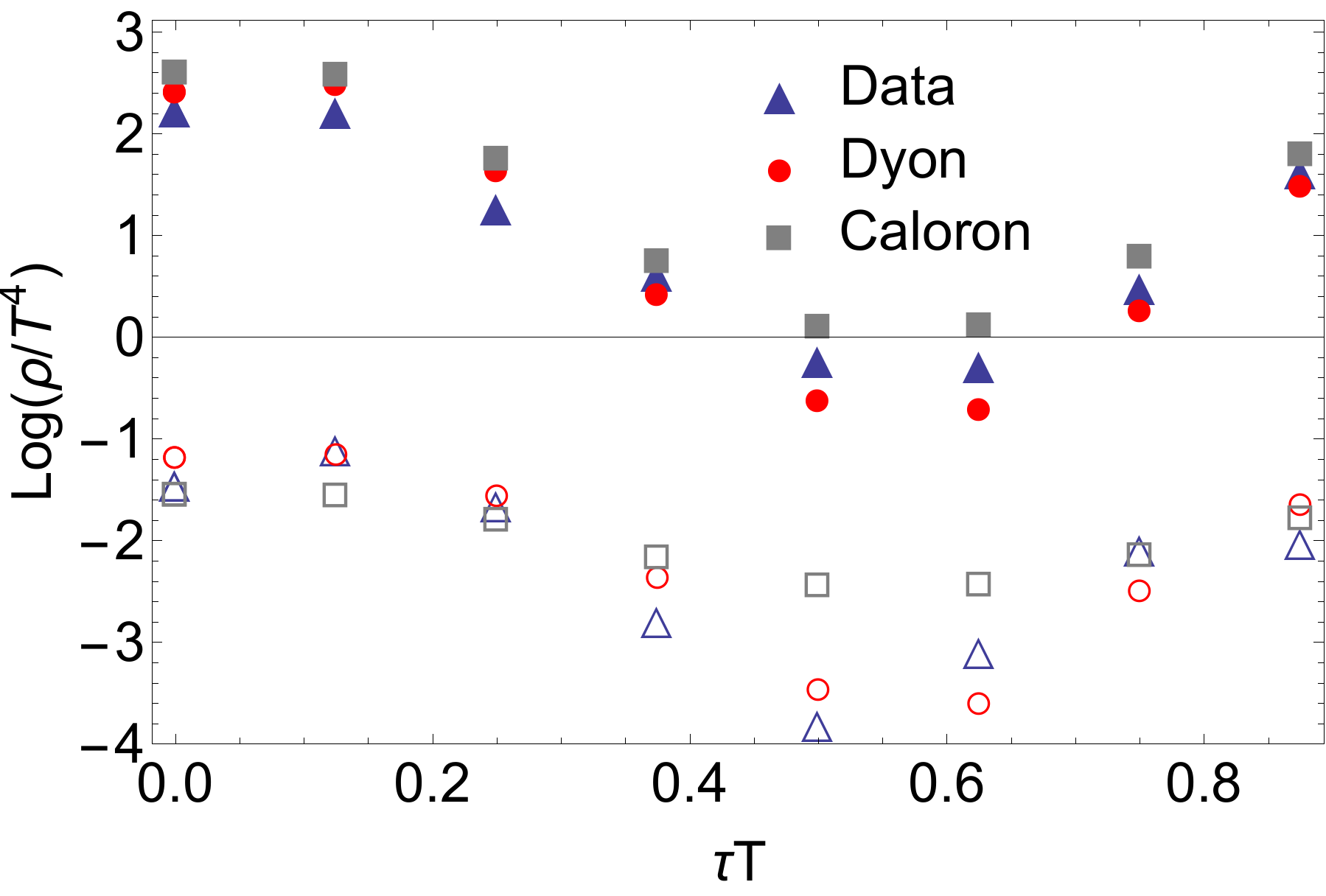}
\caption{The $\log(\rho(\tau)/T^4)$ corresponding to the zero mode of valence 
overlap fermions, whose profile was shown earlier in the middle panel of Figure~\ref{fig:zmodeloc}, for a temporal periodicity phase $\phi = \pi $. The lattice 
data for the fermion zero-mode density, shown as triangles is compared to the analytic 
density profile for a dyon, shown as circles, and for a caloron with trivial holonomy, 
shown as squares. The filled symbols represent the logarithm of the normalized 
zero-mode density at a value of the $x$-coordinate close its maxima whereas the 
empty symbols correspond to a value of $x$-coordinate far away from its maximum. 
The $\chi ^2$ $310, 423$ respectively when the fit to the lattice data is done 
to a dyon and to a caloron with a trivial holonomy.}
\label{fig:Logxtauprofile}
\end{center}
\end{figure}

Before we proceed to answer this question, we introduce another 
observable which distinguishes between a dyon and a caloron 
profile, allowing us to confirm unequivocally the identity of 
the zero-modes in our studies as dyons. We specifically calculate 
the profile of the zero-mode density along the $x$-$\tau$ plane, 
(obtained setting the other spatial coordinates at the maxima of the 
peak) and propose this as a diagonistic tool to distinguish between 
dyons and calorons. If we now compare the shape of the density profile 
obtained from lattice to that calculated from the analytical formula 
in Eq.~\ref{eqn:zmode1}, both for a caloron (with a trivial holonomy),
and a dyon as shown in Figure~\ref{fig:Logxtauprofile}, we find a much 
better agreement with the latter. In particular, the large distance 
fall-off of the densities serve as a very discriminating evidence for 
the presence of a dyon. Near the confining values of the holonomy, 
fermion density-profile corresponding to a dyon has a faster fall-off 
while fitting the zero-mode peak accurately, which is consistent with 
the lattice data. For calorons, however the large distance fall-off is 
more gradual which clearly shows deviations from the lattice profile. 
In both these plots the logarithm of the density is shown since it resolves this 
long-distance fall-off of the profile more accurately. Though the 
comparison and fit to lattice data is explicitly shown for one specific 
gauge configuration at $1.08~T_c$, with a temporal periodicity phase 
$\phi=\pi$ for the quarks, this trend is generically true for a majority 
of the configurations we have studied so far, just above $T_c$.

As explained in detail in the Appendices, we have numerically solved 
for the analytic formulae of fermion zero-mode densities corresponding to 
mutually non-interacting $L$ and $M$-dyons as in Eq.~\ref{eqn:zmode1}, 
matching their peak heights and positions to those of the zero-modes located 
within the lattice box,  for example, the ones showed in 
Figures~\ref{fig:zmodedeloc},~\ref{fig:zmodeloc}. The fact that 
the fits converged with a reasonably good precision, allows us  
to conclude that a semi-classical theory of weakly-interacting dyons 
can, quite accurately, explain the QCD vacuum quite well near 
$T_c$.

\section{Summary and Outlook}

This work is inspired by multiple earlier works spread over the 
last couple of decades,  trying to understand the underlying 
topology of the gauge fields, generated numerically in lattice 
gauge theory simulations. Determining the exact identity of the 
topological solitons is a difficult problem, requiring sufficient 
control on finite volume effects and lattice spacing artifacts. Our 
work has systematically developed more precise and discriminating tests, 
allowing us to reveal the identity of the topological objects and clusters 
present in finite temperature QCD gauge ensembles. We focused on the 
temperature interval $(1-1.2)T_c$ just above the chiral crossover transition 
temperature, $T_c$. In short, the main conclusion of this work can be 
summarized by the statement that both individual topological objects, 
and the \emph{topological clusters} observed do correspond, with unexpected 
accuracy, to zero modes associated with the instanton-dyons.

The so-called \emph{fermionic method} used in this work, allows one to probe the 
topological lumps and their interactions through the lowest eigenvalues of 
the QCD Dirac operator, instead of looking directly at (very noisy) gauge 
fields. This filtering tool has been known for quite sometime~\cite{Edwards:1999zm}, 
and has been used to show that the \emph{atoms}, i.e. the elementary building blocks 
of gauge topology, are the dyons~\cite{Gattringer:2002wh,Bornyakov:2015xao}, which 
have fractional topological charges. While these previous works could detect 
well-separated dyons, the exact identity of the so-called \emph{topological clusters} 
could not be resolved till now.

Methodical improvements we have achieved are based on two pillars. 
One is the extensive use of overlap fermions which has exact chiral 
symmetry to very high precision on finite lattices. The other is the use 
of QCD configurations generated using domain wall fermion 
discretization with physical quark masses, provided to us by the HotQCD 
collaboration. It is important that even domain wall discretization 
was found to be not so accurate for our purposes. To get the purity 
of chiral and topological properties, we need to apply the overlap 
fermions, which was a significant change on the algorithm part of the 
project.

Furthermore, we took a challenging task of devising a fitting 
procedure, to understand the constituents of~\emph{topological clusters} 
seen in the lattice simulations.  QCD with $N_c=3$ gauge group has three 
different species of dyons. With all three dyon species overlapping with each other, 
the shape of each fermion zero-mode configuration depends on twelve collective 
coordinates. The lattice data however provided a few thousands of relevant entries, 
good enough to get stable fits. The fits thus provided very nice explanation of the \emph{topological clusters} in terms of the overlapping dyons. We found that the 
corresponding topological structures sometimes consist of three well separated 
dyons, and sometimes arise from strongly overlapping ones. This qualitative finding 
is already important, as it rules out a mean-field type picture of 
the~\emph{dyon plasma}. The dyons are strongly correlated, not randomly placed.  
The fits turned out to be so good, that we could even reproduce the shape of the 
zero-modes in terms of a semi-classical theory of dyons. We further checked the 
robustness of our fitting procedure and found that the average Polyakov loop values 
constructed out of the final holonomies, extracted from the fit, agrees remarkably 
well with the corresponding values in QCD.
The other advancement we made was a very precise measurement of the near-zero 
modes. Scrutiny of near-zero modes revealed qualitatively very different 
interactions between $L$-$M$ dyons compared to that between $M$-dyons. Not only do
we clearly see that there are more $M$-type dyons than $L$-type ones, our results for
the Dirac spectral density (Figure~\ref{fig:nearzerob1771}) show that while the latter 
form finite neutral clusters, i.e., closely spaced $L\bar L$ pairs and have zero quark condensate, the $M$-dyons still do break chiral symmetry up to $1.2~T_c$.

The conclusion that emerges out of our study is highly nontrivial: the topological lumps
we study here, are located within a quark-gluon plasma, at temperatures above $T_c$, with a 
large density of thermal quarks and gluons around. There is no doubt that the gauge fields 
of these lumps are not close to classical solutions (minima of their action) but strongly 
deformed by plasma excitations. And yet, the corresponding Dirac zero eigenstates, in number 
and even exact shape, appear completely unperturbed by them, in accurate correspondence 
to what one gets from solving the Dirac equation for weakly-overlapping classical dyons. 
This conclusion is also practically relevant; if this phenomenon survives further scrutiny
near the continuum limit, it would imply that one can use semi-classical ensembles 
without thermal corrections, at least in the fermionic sector, to study topological 
properties in QCD.

Needless to say, one can improve the statistics and extend the temperature range of this 
study, finding in particular, the dyon densities at all temperatures, and elucidating 
their exact role in the formation of quark condensates. One can also deform QCD by using nontrivial flavor-dependent periodicity phases $\phi_f$ (or using imaginary 
chemical potentials) in simulations and study the intimate connection between 
deconfinement and chiral phase transitions. 
Another challenge is to get a better understanding of the effective interactions 
between the dyons. Since dyons have magnetic charges, one may wonder if the magnetic 
confinement, known to be present above $T_c$, have a role in it. In other words, it 
will be interesting to investigate how dyon-ensembles at high temperatures may 
interpolate to Polyakov's monopole-driven confinement in three dimensional gauge 
theories.

\section{Acknowledgments}
This work was supported in part by the Office of Science, U.S. Department of Energy,   
under Contract Nos. DE-FG-88ER40388 (E.S) and DE-SC0012704 (R.N.L) and by the Department of 
Science and Technology, Govt. of India through a Ramanujan fellowship (S.S). We thank the 
HotQCD collaboration, formerly also consisting of members from the RBC-LLNL collaboration, 
for sharing the domain wall configurations with us. S.S. would like to thank Margarita 
Garcia Perez and Antonio Gonzalez-Arroyo for many insightful comments and helpful 
suggestions on the first version of the draft.  We gratefully acknowledge the 
computational facilities at the Institute of Mathematical Sciences. Our GPU code 
is in part based on some of the publicly available QUDA libraries~\cite{Clark:2009wm}.

\begin{appendix}
\section{Details of the fitting procedure} 
\begin{figure}[]
\begin{center}
\includegraphics[width=7cm]{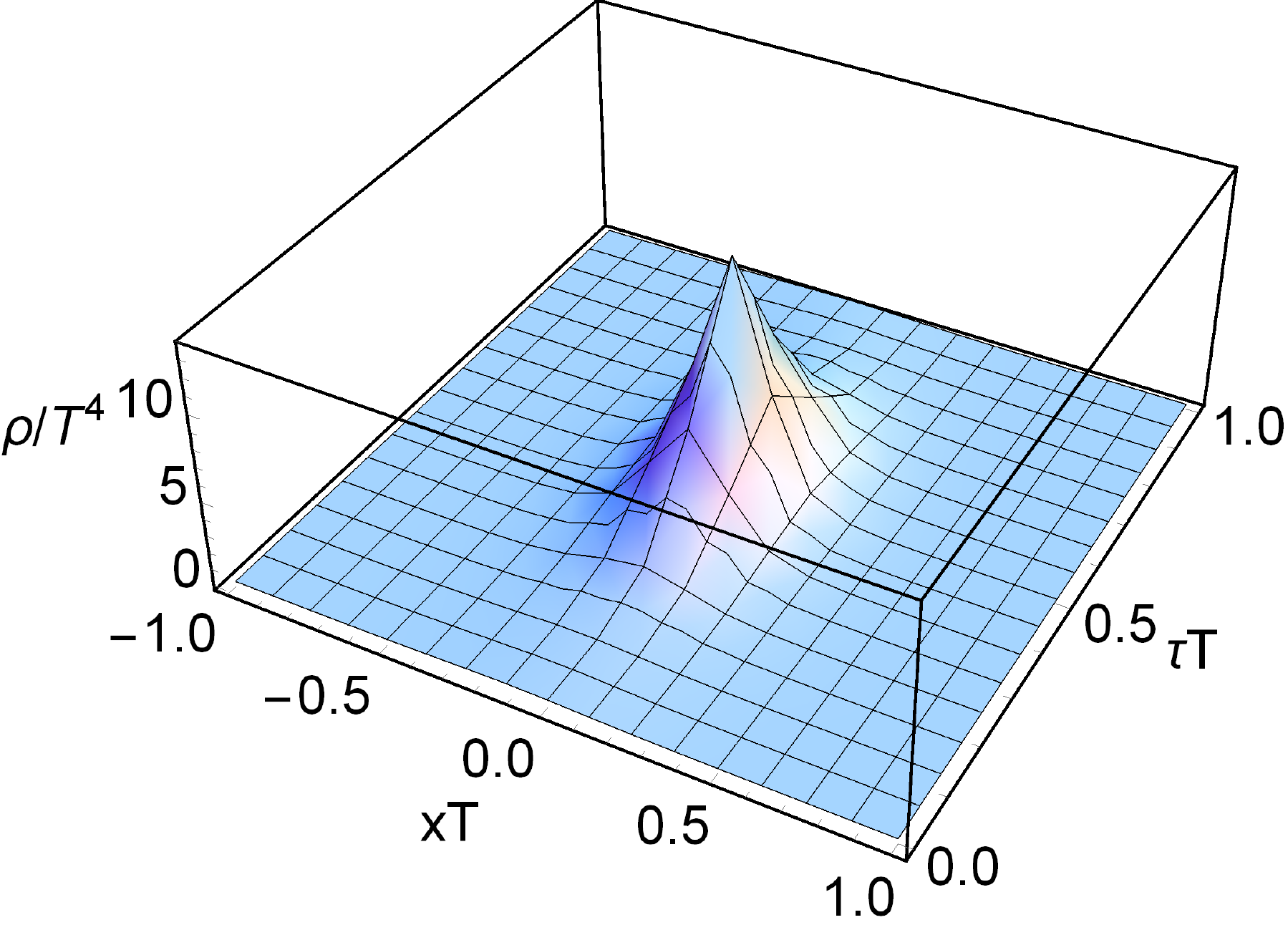}
\includegraphics[width=7cm]{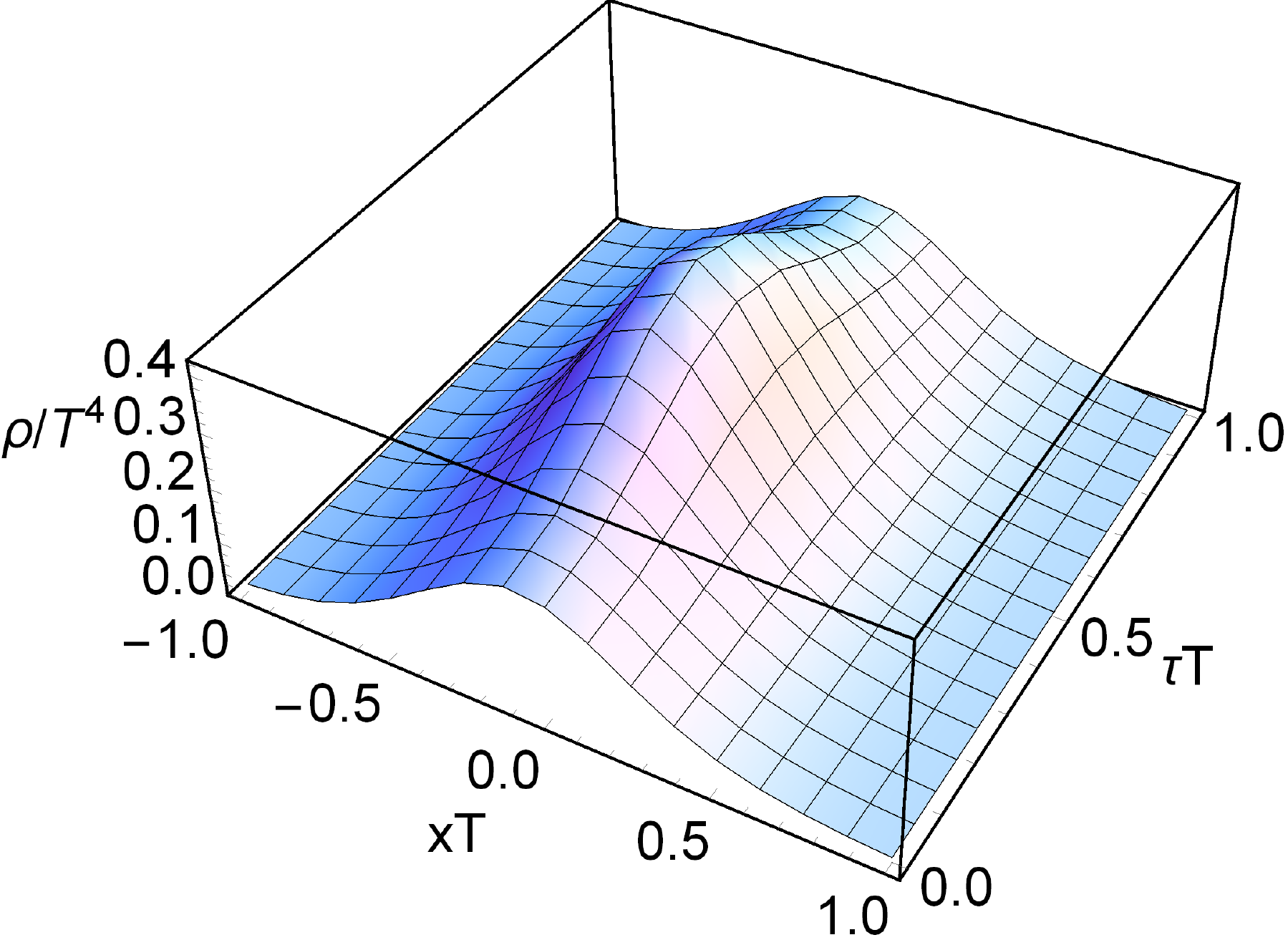}
\caption{Examples of extremes of fermion zero-mode density obtained from the analytic formula given in Eq. \ref{eqn:zmode1}. The profile of a $L$-dyon is shown for two different cases, when all 3 species of dyons in the box, $L$, $M_1$ and $M_2$ are close to each other (top panel) and secondly when all 3 dyons are far apart from each other (bottom panel). The $y$ and $z$ coordinates are chosen to be at the maxima of the zero-mode profile.}
\label{fig:Fitexamples}
\end{center}
\end{figure}

In this section, we explain the details of how we have numerically fitted the solution of a dyon density profile obtained from Eq.~\ref{eqn:zmode1} to the density of the fermion zero eigenmodes, measured on gauge configurations on the lattice. The details of how we obtain the analytic solutions corresponding to Eq.~\ref{eqn:zmode1} are given in the Appendix B. We chose the (overlap) fermion zero-mode solutions on gauge configurations with a net topological charge $\vert Q \vert =1$, for the fit.  The fit was performed simply through a gradient flow minimization of the $\chi ^2$ function, made possible thanks to a large number of data points available from our lattice calculations. However in this procedure, there is a finite probability that the fit procedure may result in ending up in a local minimum of the $\chi^2$ landscape as a function of the fit parameters. In order to circumvent this possibility, we made an initial guess for a couple of parameters based on the properties of the zero mode. For example, we set the temporal periodicity phase and the spatial co-ordinates of the dyon to be closest to the maximum or the peak of the fermion zero-mode density under study, such that the initial shape of the analytic solution mimics quite well the overlap zero-mode profile. In Figure~\ref{fig:Fitexamples}, we show typical zero-mode density profiles corresponding to a L-dyon as a function of any two space coordinates, which is simply a numerical solution of Eq.~\ref{eqn:zmode1} with the two other M-dyons either placed close or far from each other as well as the L-dyon. As evident from the Figure, the analytic dyon profiles are distinctly different depending on how close or far apart the three dyons are in a box, and hence choosing an initial trial profile as close to the realistic profiles obtained from lattice already improves the final fit. 

In total we had thirteen parameters for the fit, nine of them being the spatial coordinates 
of three dyons respectively, two of them the free angles or the holonomies $\mu_i$, the temporal periodicity phase and the normalization (height) parameter. Even the spacetime 
profile for the zero mode on $\vert Q \vert = 1$ configurations does not always comprise of a single distinct peak but may have additional smaller peaks often far enough away from each other. Thus for the analytic fermion zero-mode solution to be a good fit, we therefore considered the height of the peak as a fit parameter that could take any value between zero and one, to take into account that the peak only caries a fraction of the total normalization.  The error in estimating the zero-mode density at each spacetime point on the lattice was assigned to be identically equal, which implies that the points near the maxima of the density profiles are relatively important in deciding the convergence of the fits, since all solutions approach zero values at large distances, albeit with a different dependence on the coordinates. We found that the initial choice, in which we construct all three dyons starting from Eq. (\ref{eqn:zmode1}), close to each other such that the relative separation is $\sim 0.05$ fm (like shown in Figure~\ref{fig:Fitexamples}), is the most optimal. This is due to the fact that during $\chi ^2$ minimization, it is more likely for the dyons to move further away from each other.  Furthermore, the fits were performed with two different initial choices of average Polyakov loop $\langle P \rangle=0$, which correspond to the choice of the holonomies $\mu _1=0$, $\mu _2=2 \pi /3$, $\mu _3=4 \pi /3$ and $\langle P \rangle=1/3$  for which the holonomies are $\mu _1=0$, $\mu _2=\pi /2$, $\mu _3=3 \pi /2$. The choice of $\langle P \rangle=0$ is closest to the corresponding values physical values just above $T_c$, however $\langle P \rangle=1/3$ was chosen as an additional initial guess to check how our fit procedure performed when we are sufficiently far way from the anticipated physical value. It was however ensured that the values $\mu_i$ did not cross from $\pi /3$ and $-\pi /3$ corresponding to two distinct dyon solutions. As evident from Figure~\ref{fig:AvPolyakovLoop} the final results for the average Polyakov loop agree within errors, for either choice of the initial values, giving us a confidence on the robustness of our fitting procedure.

We have also checked the sensitivity of our numerical fitting procedure on different initial choices of the positions of dyons. The final average values of the Polyakov loop, obtained as a result of the fit, showed little dependence on the initial separation of dyons. However the final average distance between dyons resulting from the convergence of the fit did show a mild dependence on the initial separations. For different choices of the initial positions, we obtained the final separations in the range $0.3$-$0.4$ fm. This is related to the fact that a fermion zero-mode solution for a particular dyon is most sensitive to position of the  dyon situated furthest away from it. It was however assuring that the final $\chi^2$ values were not very different, giving us a confidence in our numerical procedure. This may be as well because the third dyon which is situated anywhere in between, does not affect the outcome of the final fit.

\section{Finding numerical solutions for the dyon density} 
While we did derive an analytic solution to Eq.~\ref{eqn:zmode2} for the dyon-density, the final expression turns out to be quite complicated due to the fact that eight independent boundary conditions need to be fulfilled. Hence substituting the values of the initial holonomies $\mu_i$ and the coordinates in that expression and using it as the initial trial solution of our numerical fitting algorithm was much more cumbersome. Instead, it is computationally much faster to solve the density from Eq.~\ref{eqn:zmode2} using standard numerical techniques. For this, we first solve the homogeneous part of the equation (away from the location of the delta functions), which gives us solutions of the form,
\begin{eqnarray}
f_x(\phi ) &=& C_1 \rm{e}^ { -(r - i \tau) \phi }  +  C_2 \rm{e}^ { (r + i \tau) \phi }  \nonumber
\end{eqnarray} 
where $C_1$ and $C_2$ are spacetime dependent functions but independent of $\phi$. There are four such solutions corresponding to the four regions in between the delta functions, hence we have to calculate eight independent parameters $C_1,\cdots,C_8$, which satisfy,
\begin{eqnarray}
M(x) c &=& v_0
\end{eqnarray} 
where $c$ is a column vector that contains $C_1$ to $C_8$ as its entries. The operator on the L.H.S of Eq.~\ref{eqn:zmode2} acting on $f_x$ can be represented as a  $8\times 8$ matrix  $M(x)$, containing information about the positions of the dyon zero-modes, acting on the column-vector $c$. This matrix-equation can be easily solved using linear algebra techniques giving us the solution for vector $c$. One just needs to calculate the inverse of matrix $M$ which is computationally inexpensive since it is only of size $8\times 8$. The spacetime derivatives of the vector $c$ hence follows immediately as,

\begin{eqnarray}
c &=& M^{-1} v_0~, \\
c' &=& -M^{-1} M' M^{-1} v_0~, \nonumber \\
c'' &=& -M^{-1} M ''M^{-1} v_0+2 M^{-1} M 'M^{-1}M 'M^{-1} v_0~. \nonumber
\end{eqnarray}

We recall that, in order to calculate the zero-mode density from Eq.~\ref{eqn:zmode1}, it is sufficient to calculate up to second derivative of the functions $C_i$ with respect to the spacetime coordinate $x$. The computational cost for calculating the density is mainly due to the matrix inversions, since matrix multiplications with the derivatives of $M$ are very fast. The speedup in this process of computing the zero-mode density is around 36 times as compared to using the analytic solution to Eqs~\ref{eqn:zmode1},\ref{eqn:zmode2}. This method is also highly parallizable, since the dyon zero-mode density at one position is independent of the density at any other spacetime point. We ran this code on four threads of a single processor using openMP, but it can be easily scaled to many more CPU cores.


\end{appendix}

\end{document}